\documentclass[11pt,tightenlines,eqsecnum,floats,aps,amsmath,amssymb,nofootinbib,prd,floatfix]{revtex4-2}

\usepackage{graphicx}
\usepackage{epstopdf}
\usepackage{latexsym}
\usepackage{amssymb}
\usepackage{amsmath}
\usepackage{color}
\usepackage{mathrsfs}
\usepackage{xparse}
\usepackage{float}
\usepackage{mathtools}
\usepackage{multirow}
\usepackage[center]{subfigure}
\usepackage{tensor}
\usepackage{physics}
\usepackage{amsthm}
\usepackage{ulem} 

\newcommand{\bq}{\begin{equation}}
\newcommand{\eq}{\end{equation}}
\newcommand{\nb}{\nonumber}
\newcommand{\Ncl}{N_{\rm class}}  
\newcommand{\Nacl}[1]{\ensuremath{N^{#1}_{\rm class}}} 
\newcommand{\Neff}{N}  
\newcommand{\Naeff}[1]{\ensuremath{N^{#1}}} 
\newcommand{\Ccl}{C^{\rm class}}  
\newcommand{\Cacl}[1]{\ensuremath{C_{#1}^{\rm class}}} 
\newcommand{\Ceff}{C}  
\newcommand{\Caeff}[1]{\ensuremath{C_{#1}}} 
\newcommand{\Hcancl}{\ensuremath{H_{\rm can}^{\rm class}}}  
\newcommand{\Hcaneff}{\ensuremath{H_{\rm can}}}
\newcommand{\polyF}{\ensuremath{h^j_a(A^j_a;\lambda^j_a)}}
\newcommand{\FCl}{\ensuremath{{\cal F}^{\rm class}}}
\newcommand{\Feff}{\ensuremath{{\cal F}}}
\newcommand{\FClIndex}{\ensuremath{{\cal F}^{\rm class}}}

\newtheorem{innercustomgeneric}{\customgenericname}
\providecommand{\customgenericname}{}
\newcommand{\newcustomtheorem}[2]{%
  \newenvironment{#1}[1]
  {%
   \renewcommand\customgenericname{#2}%
   \renewcommand\theinnercustomgeneric{##1}%
   \innercustomgeneric
  }
  {\endinnercustomgeneric}
}

\newcustomtheorem{customthm}{Theorem}
\newcustomtheorem{customlemma}{Lemma}
\newcustomtheorem{customcorollary}{Corollary}
\newcustomtheorem{customproof}{Proof}

\begin{document}
\newcommand{\bqn}{\begin{eqnarray}}
\newcommand{\eqn}{\end{eqnarray}}
\title{On consistent gauge fixing conditions in polymerized gravitational systems}
\author{Kristina Giesel$^{1}$}
\email{kristina.giesel@gravity.fau.de}
\author{Bao-Fei Li $^{2}$}
\email{baofeili1@lsu.edu}
\author{Parampreet Singh$^2$}
\email{psingh@lsu.edu}
\author{Stefan Andreas Weigl$^1$}
\email{stefan.weigl@gravity.fau.de}
\affiliation{$^{1}$ Institute for Quantum Gravity,  Department of Physics, Theoretical Physics III, FAU Erlangen-N\"urnberg, Staudtstr. 7, 91058 Erlangen, Germany\\
$^{2}$ Department of Physics and Astronomy, Louisiana State University, Baton Rouge, LA 70803, USA}

\begin{abstract}

For classical gravitational systems the lapse function and the shift vector are usually determined by imposing appropriate gauge fixing conditions and then demanding their preservation with respect to the dynamics generated by a canonical Hamiltonian. Effective descriptions encoding quantum geometric effects motivated by loop quantum gravity for symmetry reduced models are often captured by polymerization of connection (or related) variables in gauge fixing conditions as well as constraints. Usually, one chooses the same form of polymerization in both cases. A pertinent question is if the dynamical stability of the effective gauge fixing conditions under the effective dynamics generated by the polymerized canonical Hamiltonian is provided by the lapse function and the shift vector obtained from the polymerization of their classical counterparts. If this is the case, then we say that gauge fixing and polymerization commute. In this manuscript we investigate these issues  and obtain consistency conditions for the commutativity of gauge fixing and polymerization. Our analysis shows that such a commutativity occurs in rather special situations and reveals pitfalls in making seemingly well motivated choices which turn out to be inconsistent with the effective dynamics. We illustrate these findings via examples of symmetry reduced models in the loop quantization of the Schwarzschild interior and Lema\^{\i}tre-Tolman-Bondi (LTB) spacetimes
and report the non-commutativity of  gauge fixing  and polymerization, and inherent limitations of some choices made in the literature with a consistent effective dynamics.
\end{abstract}

\maketitle

\section{Introduction}
\label{sec:intro}

As is well-known, in order to solve canonical gravitational systems, one needs to deal with the gauge freedom encoded in the system. There exists two strategies that can be followed. In the first, one chooses appropriate gauge fixing conditions which can be used to determine the corresponding lapse and shift. Generally, requiring the stability of the chosen gauge fixing conditions leads to a set of algebraic equations that can be solved for the lapse and the shift which determines their explicit dependence on the elementary phase space variables. Then by strongly imposing the consistent gauge fixing conditions, one can work in a  space where the dynamics of the physical degrees of the freedom can be unraveled.  The second approach aims at constructing Dirac observables, which are quantities that commute with all constraints present in gravitational systems and which can be chosen as elementary variables in the reduced phase space. In the relational formalism \cite{rovelli,rovelli2} the construction of Dirac observables is strongly tied to a choice of reference fields with respect to which the dynamics of the Dirac observables is formulated \cite{Vy1994,dittrich,dittrich2, thiemann2006}. Often these reference fields are also referred to as clocks although the gauge fixing conditions might not necessarily be related to a choice of the temporal coordinate only. For certain class of gauge fixing conditions and choices of reference fields these two approaches can be related to each other, (see for instance appendix H in \cite{Giesel:2007wi}).  In literature, there are mainly two types of gauge fixing conditions involving either geometric or matter degrees of freedom respectively. In the context of the relational formalism one calls the first choice geometrical clocks and the latter matter clocks. In the models where one considers matter reference fields these fields are dynamically coupled to gravity and the gauge invariant dynamics of the remaining degrees of freedom is formulated with respect to these reference fields \cite{thiemann2006,pss2009,Giesel:2007wi,Giesel:2020raf}. Similarly, we can also choose some of the geometric degrees of freedom as the reference fields as one does in the case of vacuum gravity, (see for eg. \cite{dt2006,gh2018,ghs2018,Frob:2021ore}).

The above gauge fixing procedure in classical gravitational systems is also required and equally important for quantized systems where the quantum dynamics is prescribed by a quantum Hamiltonian operator. Here one often follows the approach where one performs a gauge fixing at the classical level yielding in general phase space dependent versions of lapse and shift and then promotes these to the corresponding operators in the quantum theory. Another strategy is to implement the gauge fixing in the quantum theory. If we formulate the two approaches in the context of the relational formalism this corresponds to choosing either classical or quantum clocks respectively. Quantum clocks in the context of the relational formalism have been discussed in various settings, see for eg. \cite{Gambini:2010ut,Gambini:2015zda,Hoehn:2019fsy,Hoehn:2020epv,Gambini:2020nsf}, where a broader perspective on quantum clocks and related results are presented than we will address in our work here.
In the quantum theory, not only lapse and shift but also the gauge fixing conditions are promoted to operators in the Hilbert space. In particular, the gauge fixing conditions now become operator relations and the preservation of these operator valued relations becomes a much more complicated problem due to the fact that we need to work with operators instead of classical quantities. An interesting question in this context is the way choosing gauge-fixing conditions or clocks either at the classical or quantum level affects the final model under consideration and under which conditions we can see a kind of commutativity of gauge-fixing and quantization in this sense. To answer this question in full loop quantum gravity (LQG) is a complicated task therefore in this work we will restrict our discussion to effective models. In those models one often uses a polymerized version of the classical theory as an effective theory which is assumed to capture the underlying loop quantization.\footnote{A  motivation for this arises from loop quantum cosmology \cite{Ashtekar:2011ni} where effective dynamics derived using coherent states can for some models be mimicked by a simple polymerization of connection (or related) components and is known to well approximate the underlying quantum dynamics \cite{Ashtekar:2006wn,Diener:2014mia,Diener:2017lde,Singh:2018rwa}. Note the usage of phrase ``effective theories" in this setting is different from the one generally employed in effective action techniques.}

In recent years, LQG techniques have been applied for various isotropic and anisotropic spacetimes in cosmology \cite{Ashtekar:2011ni}, and static and dynamical spherically symmetric spacetimes in the black hole physics (see for eg. \cite{Ashtekar:2005qt, Modesto:2005zm, Boehmer:2007ket,Campiglia:2007pb,Gambini:2008dy,Tibrewala:2012xb,Chiou:2012pg, Dadhich:2015ora, Corichi:2015xia, Yonika:2017qgo, Olmedo:2017lvt,Ashtekar:2018lag,Ashtekar:2018cay,Bodendorfer:2019cyv,Han:2020uhb,Kelly:2020uwj, Benitez:2020szx, Gambini:2020nsf, Gan:2020dkb, Giesel:2021dug,Li:2021snn}). Many of these works use the effective theory (or the polymerization) to understand Planck scale physics from LQG. As compared with the cosmological setting where on most of the occasions only several forms of the time (for example, the cosmic time and the conformal time)  are used, in the black hole spacetimes various gauges which correspond to different choices of the lapse and shift can be chosen.
Unfortunately, not much attention has been paid so far to analyze the consistency of the gauge fixing conditions as well as the choices of the lapse and shift in the effective theories for black hole and spherically symmetric spacetimes.  Previously the choices of the lapse and shift in the effective theories of the polymerized black hole were usually determined by requiring the forms of the effective constraint algebra to be the same as their classical counterparts or demanding the lapse and shift to act on the same lattice as the Hamiltonian constraint etc. In the following, we understand in detail consistency conditions to choose the lapse and shift in such effective theories, which allow us on one hand to set the guidelines for constructing consistent  models in polymerized theories and on the other hand point out problematic features with some existing models. We consider two simple example models which share features with several others. The first describes the loop quantization of the Schwarzschild interior \cite{Corichi:2015xia} and the second inhomogeneous dust collapse in LTB spacetimes \cite{Kelly:2020lec,Kelly:2020uwj}. Given the similarity of techniques which these models share with other works employing polymerization, it is quite possible that the conclusions we find in this manuscript are applicable for various other loop quantized models in black hole spacetimes.

To be specific, we aim to analyze the question of how a gauge fixing procedure can be applied in effective theories in the framework of symmetry reduced models motivated from LQG. Although we will focus in this work  on the aspect of gauge fixings to deal with gauge degrees of freedom of these models, this will also yield some insight on the comparison of classical versus quantum clocks for those cases where the gauge fixed theory can be related to the reduced phase space of the corresponding Dirac observables. For these effective models one often considers a classical gauge fixing,  applies a loop quantization, and then works with an effective theory that still has a fingerprint of the underlying quantum model. In practical applications this manifests in a polymerization of the Hamiltonian constraint and the diffeomorphism constraint respectively as well as the physical Hamiltonian, that usually has been obtained for a given choice of the classical lapse and shift. In case these also involve connection degrees of freedom the question arises how the polymerization at the effective level affects lapse and shift. Some important questions in this setting are: Do the gauge fixing conditions that are compatible with the polymerized lapse and shift encode the same type of polymerization as performed for the Hamiltonian and diffeomorphism constraints? Is there any freedom in changing the ``angles'' or ``functions'' of the polymerization for different set of variables as it is sometimes performed to simplify or enhance some features of the analysis? As we will discuss in this article the basic guiding principle here should be that lapse and shift at the effective level must  be consistent with the effective dynamics. That means that the associated gauge fixing conditions need to be preserved under the effective dynamics. Requiring this leads to strict conditions on the consistency of the dynamics vis-\'a-vis polymerization and gauge fixing conditions. This requirement is essential in the sense as it is what we need to require if we aim at implementing the gauge fixing conditions at the quantum level.

Our results show that in general polymerization does not necessarily commute with gauge fixing (Lemma \ref{lemma} and Lemma \ref{corollary3}). By this we mean that applying the same ansatz for polymerization in the classical Hamiltonian and diffeomorphism constraint as well as  for the gauge fixing conditions, the choices of lapse and shift that are consistent with the resulting effective dynamics do not in general agree with just the polymerization of their classical counterparts. Carrying this over to a more general discussion this implies that if we determine lapse and shift from a choice of a classical clock and perform their quantization, this process will result in an inconsistent  quantum dynamics if we assume that the quantum clock agrees with classical clock in the classical limit {\it{and}} it is quantized in the {\it{same}} way as the constraints\footnote{In effective theories this corresponds to using the same polymerization for the gauge fixing conditions and the constraints.}. As shown in Lemma \ref{corollary3} this non-commutativity between the gauge fixing and the polymerization generally holds for fully constrained systems under the conditions that there is at least one time-dependent gauge fixing condition which depends at most on one canonical pair for which at least one variable of the pair is polymerized. Therefore, coordinate gauge fixings that involve a temporal dependence for gravitational systems in which part of the degrees of freedom are loop quantized do often belong to this class of models.
Removing one of the above conditions can make gauge fixing commute with polymerization. For example, a consistent effective lapse and shift coincide with the polymerization of their classical counterparts if each gauge fixing condition depends at most on one and in general different canonical pair and in addition none of the gauge fixing conditions depends on the temporal coordinate (Corollary \ref{lemma2}), or the time-dependent gauge fixing condition only include unpolymerized matter degrees of freedom (Corollary \ref{lemma3}),  or the time-dependent gauge fixing condition only depends on the geometrical degrees of freedom but the polymerization is only implemented in the matter sector.

Usually when gauge fixing does not commute with polymerization,  working with the polymerization of the classical lapse and/or shift requires one to modify the gauge fixing condition appropriately to obtain a consistent dynamical system at the effective level. The consistent gauge fixing conditions for the effective dynamics can be obtained by solving a partial differential equation (PDE) derived from the stability requirement. We show explicitly how this is done in one of the given examples which deals with the choice of the gauge fixing conditions once the lapse is known in the context of the loop quantization of the interior of the Schwarzschild black hole \cite{Corichi:2015xia}.  In contrast, when gauge fixing does commute with  polymerization, the lapse and shift consistent with the effective dynamics must be chosen as the polymerization of their classical counterparts in order for the effective system to be dynamically consistent and to possess the correct classical limit (Lemma \ref{lemma4}). An example in which gauge fixing and polymerization commute is the inhomogeneous collapse of the dust cloud in LTB spacetime where the temporal gauge consists matter degrees of freedom only and the gauge fixing related with diffeomorphism constraints involves variables from geometrical sector only and is time-independent. This exercise provides an important lesson on the inconsistency of some choices made for the shift vector in the literature \cite{Kelly:2020lec} in relation to the consistency of the effective dynamics and the corresponding physical Hamiltonian (Lemma 3).

Furthermore, we also extend our analysis to matter reference models or where matter clocks are chosen. When the matter sector is not loop quantized, and if the  matter couples only via the triad variables,\footnote{An exception to this can arise using gauge-covariant fluxes result in a non-minimally coupled matter, i.e. with a coupling to connection, even if matter is not polymerized \cite{Liegener:2019ymd}.}  the matter contributions to the Hamiltonian/diffeomorphism constraint as well as to the physical Hamiltonian do not involve any polymerization effects at the effective level. As a consequence, for a set of gauge fixing conditions that involve only matter reference fields polymerization and gauge fixing commute. However, if we also loop quantize the matter sector, the gauge fixing will become non-commutable with the polymerization again for the same reason as choosing geometric clocks when the geometric sector is polymerized (Lemma \ref{corollary3}). We show this explicitly for the Brown-Kucha\v r \cite{Brown:1994py,Giesel:2007wi,Giesel:2007wn} and Gaussian dust model \cite{Kuchar:1990vy,Giesel:2012rb} well as the four scalar reference field model \cite{Giesel:2016gxq}.

The paper is structured as follows. In Sec. \ref{sec:GenProof}, we first present a proof of Lemma \ref{lemma} for generic geometrical clocks which shows that for this type of gauge fixing conditions polymerization and gauge fixing will not commute as long as at least one of the gauge fixing conditions is time-dependent. We then extend our analysis to several other situations when some restrictions on the assumptions for  Lemma \ref{lemma} are changed/ a bit relaxed. These analyses lead to Corollary \ref{lemma2} and \ref{lemma3}. With Lemma \ref{corollary3} we show that for a wide range of models that use a temporal gauge fixing condition involving canonical variables where at least one of the elementary variables is polymerized,  gauge fixing and polymerization do not commute. Further we analyze in Lemma \ref{lemma4} on the possibility of reverse-engineering the set of gauge fixing conditions for a given lapse and shift. In Sec. \ref{sec:Examples}, we discuss several models known from  literature as examples to clarify the results of our lemmas and corollaries. These models include vacuum and dynamical black hole spacetimes in symmetry reduced models of LQG as well as the Brown-Kucha\v r  dust model. We explicitly analyze whether the choices of the gauge fixing conditions with the polymerized lapse and/or shift in these models lead to a consistent dynamics at the effective level. Besides, the different properties of geometrical and matter clocks in the context of gauge fixing in the effective theory are addressed in the Brown-Kucha\v r dust model with a particular focus on how the situation changes if we also apply a polymerization to the matter sector. In Appendix \ref{appendix} we present a similar analysis for the Gaussian dust model and the 4-scalar field model. Finally, in Sec. \ref{sec:Conclusion} we summarize our main results.

\section{General Analysis on the consistency of gauge fixings in polymerized models}
\label{sec:GenProof}

In this section we present some proofs which provide insights on the process of polymerization in effective theories of gravitational systems based on  LQG and gauge fixing conditions. We start with the underlying assumptions which lead us to Lemma 1 which shows that when a gauge fixing condition involves an explicit dependence on the temporal coordinate as well as on the geometric degrees freedom, then polymerization and gauge fixing do not commute. This is followed by Corollaries detailing some special cases allowing for commutability of gauge fixing and polymerization. On the other hand Lemma \ref{corollary3} in a sense generalizes foregone Lemma \ref{lemma} to a broader class of gauge fixing conditions that can for instance also involve polymerized matter degrees of freedom and as shown gauge fixing and polymerization do not commute in this case. Next we discuss how in principle one could reverse engineer the gauge fixing conditions for given lapse and shift. We conclude with a Lemma 3 emphasizing for certain systems the need of exactly the same polymerization in lapse and shift as the constraints and the physical Hamiltonian to avoid any inconsistencies in dynamics. These Lemmas are applied in the next section to models on loop quantization of Schwarzschild interior \cite{Corichi:2015xia} and inhomogeneous dust collapse in LTB spacetimes \cite{Kelly:2020lec} highlighting various issues related to gauge fixing with their construction. In this section, we make a  brief comment on the model in \cite{Gambini:2020nsf} which fits Corollary 1.

Since only for very specific models and choices of gauge fixing conditions, polymerization and gauge fixing commute we choose a class of gauge fixing conditions which while are not the most general ones but are sufficiently general to cover all the examples discussed in Sec. \ref{sec:Examples} and also to demonstrate the main reason why gauge fixing and polymerization do not commute or commute in special cases. Furthermore, because we discuss examples with spherically symmetric as well as Kantowski-Sachs models we perform the proof not for specific chosen models but keep it at a more general level providing us with a better intuition on what kind of properties are characteristic of a given model. To formulate the effective model we will assume that a subset of variables is replaced by their corresponding polymerizations and all proofs in this section apply only to those models where the corresponding effective model is of this kind. These polymerizations are motivated from the full theory where holonomies play the role of the elementary variables and they involve integrals of the connection along the edges of a graph associated with a given spin network function. If there is a relation between the full theory and the effective model, then the polymerization parameter, denoted below by $\lambda$'s, needs to encode detailed properties of the underlying dynamics in the full theory. Therefore, there might exist effective models that have a more complex structure as far as their polymerization is considered than we will consider here. Nevertheless, a wide range of symmetry reduced effective models in the literature particularly for spherically symmetric or cosmological spacetimes can be characterized by the type of polymerization we consider in this work here (see also footnote 1).

To start the discussion of Lemma 1, for the polymerization as well as the class of gauge fixing conditions we assume them to have the following properties:
\begin{itemize}
    \item[1.] We only consider gauge fixing conditions for the Hamiltonian and diffeomorphism constraints.
    \item[2.]
    Furthermore, we assume that all gauge fixing conditions weakly commute with the Gauss constraints. This can for instance be achieved if we work with gauge invariant variables with respect to the Gauss constraint.
    \item[3.] We denote the set of gauge fixing conditions by $\{G_I\}$ with $I=0, \cdots, 3$. For at least one of the gauge fixing conditions $G_I$ we have $\frac{\partial G_I}{\partial t}\not\approx 0 $.
    \item[4.] All gauge fixing conditions depend on the gravitational degrees of freedom only.
    \item[5.] The polymerization of the connection variables denoted by $A^j_a$ is performed by $A^j_a\mapsto\polyF$ in the constraints and gauge fixing conditions where $\polyF$ does not depend on partial derivatives of $A^j_a$ and is not the identity function. The triad variables $E^a_j$ are not polymerized at the effective level.
\end{itemize}

Let us briefly comment on these assumptions. The first assumption is motivated by the fact that the Gauss constraint involved in formulation of general relativity (GR) in terms of Ashtekar variables is often solved in the quantum theory and then there exists no additional equations stemming from a stability requirement of the gauge fixing conditions from the Gauss constraint.  The second assumption has the consequence that the Lagrange multiplier coming from the Gauss constraint does not contribute to the stability equations and this is the situation that occurs in the models discussed in this work here. Assumption three ensures that we can work with a coordinate gauge fixing that involves the temporal coordinate and the fourth assumption restricts the choice of gauge fixing conditions to the class of geometrical clocks.
The fifth assumption allows to consider all polymerization functions that have been used in the literature so far and this allows to formulate our result independently of the specific choice of polymerization.

Furthermore, we have considered in the assumptions a polymerization of the spatial diffeomorphism constraint at the effective level as well. There exist effective models as for instance in \cite{Han:2020uhb} where only the Hamiltonian constraint is polymerized. In case we remove the polymerization of the spatial diffeomorphism constraint in the proof, the final conclusion that gauge fixing and polymerization will not commute will still hold. Because an operator for the spatial diffeomorphism constraint operator does not exist in full LQG, the way we can obtain the form of a polymerized spatial diffeomorphism constraint works as follows: We can implement an operator of the classical expression $q^{ab}C_aC_b$  expressed in terms of connection and triad variables along the lines of \cite{Giesel:2007wn}. Then we can compute its semiclassical expectation value leading in lowest order to a polymerized version of this quantity from which we can read off the polymerized form of the spatial diffeomorphism constraint. At the effective level we assume that this agrees with the classical spatial diffeomorphism constraint if we replace the connection variables by their corresponding polymerized quantities.

Now we proceed to prove the following lemma:
\begin{customlemma}{1}
\label{lemma}

Given the assumptions 1.-5. stated above then gauge fixing and polymerization will not commute. This means that the lapse function and shift vector that are consistent with the effective dynamics do not agree with the polymerization of the classical lapse function and classical shift vector.
\end{customlemma}
\begin{customproof}{of lemma 1}
~\\
\normalfont
Without any loss of generality we choose $G_0$, namely the gauge fixing condition for the Hamiltonian constraint, to be the one that depends on the temporal coordinate. In the literature so-called coordinate gauge fixings, i.e. $G^{\rm class}_I(A(t,x),E^a_j(t,x);t) = f_I(A,E) - \tau_I(t,x)$, are a popular class of gauge fixings, but it is not much more difficult to prove the lemma for general gauge fixing conditions and so we restrain from using this simplification. The elementary Poisson bracket for the gravitational degrees of freedom is given by
\bq
\{A^j_a(x,t), E^b_k(y,t)\} = \frac{\kappa}{2}\delta(x,y)\delta^j_k\delta^b_a,
\eq
where $\kappa=16\pi G$ and $G$ is the Newton's constant. All the other degrees of freedom will be labeled by $\{\Phi^A, \pi_A\}$, where the index $A$ is a multi index such that spinor valued fields are allowed as well.
At the classical level the stability requirement reads
\begin{eqnarray}
\label{eq:StabGR}
\frac{\dd}{\dd t}G^{\rm class}_I(t,x) &=& \{G^{\rm class}_I(t,x), \Hcancl \} +\delta_{I,0}\frac{\partial G_0}{\partial t}(t,x) \nonumber \\
&\approx&  \int\limits d^3y \{G^{\rm class}_I(t,x), \left(\Ncl \Ccl +\Nacl{a}\Cacl{a}\right)(y)\}
+\delta_{I,0}\frac{\partial G_0}{\partial t}(t,x)
\approx 0,
\end{eqnarray}
where we have denoted all classical quantities in particular the lapse and the shift with a corresponding label and used the second assumption that all $G_I$'s weakly commute with the Gauss constraint. The weak equality in the last step involves all constraints as well as all gauge fixing conditions. The condition in \eqref{eq:StabGR} yields the following system of equations
\begin{eqnarray}\label{eq:stabilityeq_class}
\int &d^3y& \int d^3z\,\Ncl^J \FCl_{\Neff^J,I}(x,y,z) + \delta_{I,0}\frac{\partial G_0}{\partial t}(t,x)\approx  0, \quad I=0,\cdots 3.
\end{eqnarray}
Here we have introduced the following abbreviations
\begin{eqnarray*}
\FCl_{\Neff^J,I}(x,y,z)&:=& \frac{\delta G^{\rm class}_I(t,x)}{\delta A^k_b(z)}\frac{\delta \Cacl{J} (y)}{\delta E^b_k(z)} - \frac{\delta G^{\rm class}_I(t,x)}{\delta E_k^b(z)}\frac{\delta \Cacl{J}(y)}{\delta A_b^k(z)}\\
&=&\left(\FClIndex_{I,A}\right)^b_k(x,z)\left(\FClIndex_{\Neff^J,E^a_j}\right)^k_b(y,z)-\left(\FClIndex_{I,E^a_j}\right)^b_k(x,z)\left(\FClIndex_{\Neff^J,A}\right)^k_b(y,z),
\end{eqnarray*}
as well as the compact notation $\Ncl^J = \qty(\Ncl,\Nacl{a})$ and $\Cacl{J} = \qty(\Ccl, \Cacl{a})$.
If we perform the gauge fixing directly at the level of the effective dynamics then we introduce the effective Hamiltonian and diffeomorphism constraints denoted by $\Caeff{J} = \qty(\Ceff, \Caeff{a})$ respectively and given by
\begin{equation}
\label{eq:Conseff}
\Caeff{J}=\Cacl{J}(\polyF,E^a_j,\Phi^A, \pi_A),
\end{equation}
where $\Phi^A,\pi_A$ denote all remaining degrees of freedom but the gravitational ones and $\polyF$ denotes a generic polymerization function.
Given these we replace $\Hcancl(A^j_a,E^a_j,\Phi^A, \pi_A)$ by its effective analogue given by
\begin{equation}
\Hcaneff:=\Hcancl(\polyF,E^a_j,\Phi^A, \pi_A),
\end{equation}
and furthermore introduce the effective gauge fixing conditions
\begin{equation}
\label{eq:GFeff}
G_I=G_I^{\rm class}(\polyF, E^a_j),
\end{equation}
where we have used the fact that by assumption four gauge fixing conditions $G_I's$ do depend on $A,E$ only. The stability requirement for the effective gauge fixing conditions has the form
\begin{eqnarray}
\label{eq:StabEff}
\frac{\dd}{\dd t}G_I(t,x) &=& \{G_I(t,x), \Hcaneff \} +\delta_{I,0}\frac{\partial G_0}{\partial t}(t,x) \nonumber \\
&\approx&  \int\limits d^3y \{G_I(t,x), \left(\Naeff{J}\Caeff{J}\right)(y)\}
+\delta_{I,0}\frac{\partial G_0}{\partial t}(t,x)
\approx 0,
\end{eqnarray}
which, similar to the situation in the classical theory, can be rewritten as
\begin{eqnarray}\label{eq:stabilityeq_eff}
 \int d^3y \int d^3z \,\Naeff{J}\Feff_{\Naeff{J},I}(x,y,z)+\delta_{I,0}\frac{\partial G_0}{\partial t}(t,x)&\approx & 0 ,\quad I=0,\cdots 3.
\end{eqnarray}
In the case of the effective theory the involved $\Feff_{\Neff^J,I}$ are given by
\begin{eqnarray*}
 \Feff_{\Neff^J,I}(x,y,z):= \frac{\delta G_I(t,x)}{\delta A^k_b(z)}\frac{\delta \Caeff{J} (y)}{\delta E^b_k(z)} - \frac{\delta G_I(t,x)}{\delta E_k^b(z)}\frac{\delta \Caeff{J}(y)}{\delta A_b^k(z)}.
\end{eqnarray*}
Now in case that polymerization and gauge fixing commute in the sense introduced above we need to have
\begin{eqnarray}
\label{eq:CondCommute}
  \Feff_{\Neff^J,0}(x,y,z)&=&   \FCl_{\Neff^J,0}[\polyF,E^a_j,\Phi^A, \pi_A](x,y,z)\quad{\rm and} \nonumber\\
   \Feff_{\Neff^J,a}(x,y,z) &=&  \FCl_{\Neff^J,a}[\polyF,E^a_j,\Phi^A, \pi_A](x,y,z){\cal H}(A^j_a),
\end{eqnarray}
since then the system of equations in \ref{eq:stabilityeq_class} and \ref{eq:stabilityeq_eff} have the same solution for lapse and shift (modulo polymerization).
By this notation we mean that $\Feff_{\Naeff{J},I}$ agree with the quantities we obtain by simply polymerizing the classical $\FCl_{\Naeff{J},I}$. For the  $\Feff_{\Naeff{J},a}$ we can slightly relax the condition because the gauge fixing conditions associated with $G_a$ carry no explicit time dependence and therefore we can allow an additional generic function ${\cal H}(A)$ depending on the connection variables on the right hand side. This is possible due to the absence of the term involving $\frac{\partial G_0}{\partial t}$ in the stability equation, which allows us to factor out the term involving ${\cal H}(A)$.  We now show that under the assumptions 1.-5. listed above these conditions in \eqref{eq:CondCommute} are never satisfied. For this purpose we aim at expressing $\Feff_{\Naeff{J},I}$  in terms of $\FCl_{\Naeff{J},I}$. Using the explicit form of the effective gauge fixing conditions and constraints from \eqref{eq:GFeff} and \eqref{eq:Conseff} respectively, we obtain
\begin{eqnarray}
\frac{\delta G_I(t,x)}{\delta A^k_b(z)} &=&
\int d^3z^\prime \frac{\delta G_I(t,x)}{\delta h^\ell_c(z^\prime)} \frac{\delta h^\ell_c(z^\prime)}{\delta A^k_b(z)}
=\int d^3z^\prime \frac{\delta G^{\rm class}_I}{\delta A^\ell_c(z^\prime)}[h^\ell_c(A^\ell_c;\lambda^\ell_c),E^a_j](t,x) \frac{\delta h^\ell_c(z^\prime)}{\delta A^k_b(z)},\nonumber \\
\frac{\delta \Caeff{J}(t,y)}{\delta A^k_b(z)} &=&
\int d^3z^\prime \frac{\delta \Caeff{J}(t,y)}{\delta h^\ell_c(z^\prime)} \frac{\delta h^\ell_c(z^\prime)}{\delta A^k_b(z)}
=\int d^3z^\prime \frac{\delta \Cacl{J}}{\delta A^\ell_c(z^\prime)}[h^\ell_c(A^\ell_c;\lambda^\ell_c),E^a_j,\Phi^A, \pi_A](t,y) \frac{\delta h^\ell_c(z^\prime)}{\delta A^k_b(z)}\nb.\\
\end{eqnarray}
The notation in the last in each equation means that we take the functional derivative of the classical gauge fixing conditions, take the result and replace all the remaining connection variables by their polymerized counterparts and we have used the fifth assumption on the polymerization here. If we further take into account that
\begin{eqnarray}
\frac{\delta G_I(t,x)}{\delta E_k^b(z)} &=&\frac{\delta G^{\rm class}_I}{\delta E_k^b(z)}[h^j_a(A^j_a;\lambda^j_a),E^a_j](t,x),\nonumber\\
\frac{\delta \Caeff{J}(t,y)}{\delta E_k^b(z)} &=&\frac{\delta \Cacl{J}}{\delta E_k^b(z)}[h^j_a(A^j_a;\lambda^j_a),E^a_j,\Phi^A, \pi_A](y),
\end{eqnarray}
then we obtain
\begin{eqnarray}\label{function1}
  \Feff_{\Neff^J,I}&= &  \int d^3z^\prime \Big[
\left(\FClIndex_{I,A}\right)^c_\ell [h^j_a(A^j_a;\lambda^j_a),E^a_j,\Phi^A, \pi_A](x,z^\prime)
\left(\FClIndex_{\Neff^J,E}\right)^k_b[h^j_a(A^j_a;\lambda^j_a),E^a_j,\Phi^A, \pi_A](z,y) \nonumber \\
 &&- \left(\FClIndex_{I,E}\right)^k_b [h^j_a(A^j_a;\lambda^j_a),E^a_j,\Phi^A, \pi_A](x,z)
\left(\FClIndex_{\Neff^J,A}\right)^c_\ell[h^j_a(A^j_a;\lambda^j_a),E^a_j,\Phi^A, \pi_A](z^\prime,y)
  \Big] \frac{\delta h^\ell_c(z^\prime)}{\delta A^k_b(z)} \nonumber\\
  &=&
  \Big[
\left(\FClIndex_{I,A}\right)^c_\ell [h^j_a(A^j_a;\lambda^j_a),E^a_j,\Phi^A, \pi_A](x,z)
\left(\FClIndex_{\Neff^J,E}\right)^k_b[h^j_a(A^j_a;\lambda^j_a),E^a_j,\Phi^A, \pi_A](z,y) \nonumber \\
 &&- \left(\FClIndex_{I,E}\right)^k_b [h^j_a(A^j_a;\lambda^j_a),E^a_j,\Phi^A, \pi_A](x,z)
\left(\FClIndex_{\Neff^J,A}\right)^c_\ell[h^j_a(A^j_a;\lambda^j_a),E^a_j,\Phi^A, \pi_A](z,y)
  \Big] {\cal H}^{\ell b}_{ck}(A). \nonumber\\
\end{eqnarray}
Here in the final step we have used assumption 5, namely that $\polyF$ does not depend on the derivatives of $A^j_a$ and hence the functional derivative that is factored out in the one before the last step involves $\delta(z^\prime, z)$ linearly. Further we introduce the following notation
\begin{equation}
    \frac{\delta h^\ell_c(z^\prime)}{\delta A^k_b(z)} = {\cal H}^{\ell b}_{ck}(A) \delta(z^\prime, z).
\end{equation}

Since by assumption five the polymerization function $\polyF$ cannot be the identity map the most simple form the function ${\cal H}^{\ell b}_{ck}(A)$ can have is
\begin{equation}
\label{polymerizationsimple}
{\cal H}^{\ell b}_{ck}(A) = {\cal H}(A^\ell_c)\delta^{b}_{c}\delta^\ell_{k}   \quad {\cal H}\not={\rm id}.
\end{equation}
If we reconsider the condition written in \eqref{eq:CondCommute} we realize that although it is satisfied for $ \Feff_{\Naeff{J},a}(x,y,z)$ when all spatial gauge-fixing conditions $G_a$  depend on at most one and in general different canonical pair only, it can never be satisfied for  $\Feff_{\Naeff{J},0}(x,y,z)$ unless ${\cal H}(A^c_\ell)$ is the identity map which is excluded by our assumption. Hence, we have shown that for the class of gauge fixing conditions that satisfy assumptions 1-5 polymerization and gauge fixing will not commute. $\qedsymbol$
\end{customproof}

Before  discussing some specific examples in the next section, we want to extend this analysis to some other classes of gauge fixed models that are present in the literature. As we will see, we can extensively use the calculations of the foregoing lemma to prove the following corollaries.
First we consider the stationary slicings of the spacetime manifold. In other words we will consider models which have no temporal dependence in the gauge fixing conditions. Otherwise assuming the non-contradicting properties left that were stated before, we can prove that gauge fixing and polymerization commute for such models.
\begin{customcorollary}{1}
\label{lemma2}
~\\
Given the assumptions 1.-5. stated beforehand where assumption three is replaced by
\begin{itemize}
    \item[3.] For all gauge fixing conditions we have $\frac{\partial G_I}{\partial t}\approx 0$ and each gauge-fixing condition depends on at most one and in general different canonical pair only 
\end{itemize}
then gauge fixing and polymerization will commute.
\end{customcorollary}
\begin{customproof}{of corollary 1}
~\\
\normalfont
 Using the connection of the functions $\Feff_{\Naeff{J},I}$ to the ones of the classical theory given in equation \ref{function1}  we compute
\begin{eqnarray}
\frac{\dd}{\dd t}G_I(t,x) &=& \int d^3y \int d^3z \Naeff{J}\Feff_{\Naeff{J},I}(x,y,z)\nb\\
&= & \int d^3y \int d^3z \Big(\Naeff{J}\FCl_{\Naeff{J},I}[\polyF,E^a_j,\Phi^A, \pi_A](x,y,z)\Big){\cal H}(A^c_\ell)\nb\\
&\approx & 0.
\end{eqnarray}
Since each gauge-fixing condition depends on at most one and in general different canonical pair only we can factor out the derivative of the polymerization function ${\cal H}(A^c_\ell)$ and the remaining set of equations are exactly of the same form as in the classical theory. This concludes the proof. $\qedsymbol$
\end{customproof}
We can find in \cite{Gambini:2020nsf} a model that falls in the class of such gauge fixings. The authors use the areal gauge $G_1 = \tensor{E}{^a} - g(x)$ (we have adapted the expressions to the notation used in Sec. \ref{JorgeEd_model}). The second gauge  condition is $G_0 = b - h(x)$. A quick computation shows that for this model gauge fixing and polymerization commute. Note the similarities of this model to the LTB model discussed in Sec. \ref{JorgeEd_model}, in particular one can see this in the result for the shift vectors of the two models, which modulo the lapse function coincide.

The second class we want to consider are so-called matter clocks. This means we will use an additional matter field to write a coordinate gauge. The most common form used in literature is $G_0 = T - t$, but in the following corollary we can generalize to $G_0 \coloneqq G_0(\Phi^{A},\pi_A;t)$, where $\Phi^{A}$ are additional matter fields with canonical momenta $\pi_A$. The other gauge fixing conditions remain ``geometrical", i.e. only depend on the gravitational degrees of freedom. Then the following corollary holds:
\begin{customcorollary}{2}
\label{lemma3}
~\\
Given the assumptions 1.-5. stated beforehand where we replace assumption three and four by
\begin{itemize}
    \item[3.] Only gauge fixing condition $G_0$ has a weakly non-vanishing dependence on the temporal coordinate and further matter degrees of freedom, thus we can write $G_0 \coloneqq G_0(\Phi^{A},\pi_A;t)$. The remaining gauge fixing conditions are functions of at most one and in general different canonical pair of the gravitational degrees of freedom only,
\end{itemize}
then gauge fixing and polymerization will commute.
\end{customcorollary}
\begin{customproof}{of corollary 2}
~\\
\normalfont
 Note that the stability equation of $G_0$ at the effective level is the same equation that one obtains at the classical level up to that $\Ncl, \Nacl{a}$  are replaced by $\Neff,\Naeff{a}$ since
 \begin{equation}
\frac{\dd}{\dd t}G_0(t,x) = \{G_0,\Hcaneff\} + \frac{\partial G_0}{\partial t} = \{ G^{\rm class}_0, H^{\rm class}_{\rm mat} \} + \frac{\partial}{\partial t} G_0^{\rm class}\approx 0,
 \end{equation}
 where $H_{\text{mat}}^{\rm class}$ denotes the matter contribution to $\Hcaneff^{\rm class}$ and by construction the classical expression does not involve any polymerized variables. The reason why only $\Hcaneff^{\rm class}$ contributes to above equation is that the matter contribution depends on the densitized triads and matter degrees of freedom only.
 Considering the remaining gauge fixing conditions we can make the same computation as in Lemma \ref{lemma2}. We see after factoring out the polymerization factor ${\cal H}(A^c_\ell)$ in the stability equations, the equations are the same as the classical ones. $\qedsymbol$
\end{customproof}

From the previous lemma and corollaries we learn that the temporal dependence in the gauge fixing conditions plays a pivotal role in answering the question whether gauge fixing and polymerization commute. Lemma \ref{lemma} and Corollary \ref{lemma2} show that when using a suitable type of polymerization additional contributions in the gauge-fixing conditions that do not carry any temporal dependence can be completely factored out. Hence, these stability equations associated with these non-temporal gauge fixing conditions are equivalent to their classical counterparts up to polymerizations of the involved variables in the solutions of lapse and shift. As a consequence, discussed in Corollary \ref{lemma2}, if none of the gauge fixing conditions depends on the temporal coordinate gauge fixing and polymerization commute. As can be seen from Lemma \ref{lemma} and Corollary \ref{lemma3} whether gauge fixing and polymerization commute if at least one of the gauge fixing conditions has a temporal dependence depends on the fact whether the variables involved in the gauge fixing conditions belong to pairs of canonical variables for which at least one of the variables is polymerized. Since Lemma \ref{lemma} and Corollary \ref{lemma3} focus on geometrical clocks and matter clocks respectively we can generalize our analysis to more generic gauge fixing conditions and constrained systems and  show that if
variables are involved in a temporal gauge fixing condition that are polymerized or that are not polymerized but whose conjugate variable is polymerized this generally breaks the commutativity of polymerization and gauge fixing. In the case that we restrict to temporal gauge fixing conditions that depend at most on one canonical pair we can prove the following result:
\begin{customlemma}{2}
\label{corollary3}
~\\
\begin{itemize}
\vspace{-0.5cm}
\item[1.] Given a fully constrained classical system with canonical variables $(Q^A(t,x),P_A(t,x))$ with $A=0\cdots,K$, that is the canonical Hamiltonian of the system can be written as $H_{\rm can}=\int d^3x N^I(x)C_I(x)$ for some suitable finite range of the index $I$ and a set of gauge fixing conditions $\{G_I(t,x)\}$ whose dynamically stability yields solutions $N^I_{\rm class}(t,x)$.
    \item[2.] The
set of gauge fixing conditions is such that at least one gauge fixing condition has a temporal dependence that is not weakly vanishing, i.e. $\frac{\partial G_I}{\partial t}\not\approx 0 $.
    \item[3.] At least one gauge fixing condition for which $\frac{\partial G_I}{\partial t}\not\approx 0 $  does further at most depend on one canonical pair only denoted with loss of generality by $\big(Q^0(t,x),P_0(t,x)\big)$.
  \item[4.]  The polymerization is performed according to $Q^0\mapsto h_{Q^0}(Q^0;\lambda_{Q^0}), P_0\mapsto h_{P_0}(P_0;\lambda_{P_0})$ where  $h_{Q^0}(Q^0;\lambda_{Q^0})$ and $h_{P_0}(P_0;\lambda_{P_0})$ do not depend on partial derivatives of $Q^0$ and $P_0$ respectively, at least one of the polymerization functions is not the identity function and we have $\int d^3z\Big(\frac{\partial h_{Q^0}(z^\prime)}{\partial Q^0(z)}\Big) \Big(\frac{\partial h_{P_0}(z^{\prime\prime})}{\partial P_0(z)}\Big)\not\approx \delta^{(3)}(z^\prime,z^{\prime\prime})$.
  \item[5.] All remaining variables $(Q^A,P_A)$ with $A=1,\cdots K$ can be polymerized according to $Q^A \mapsto h_{Q^A} (Q^A; \lambda_{Q^A}), P_A \mapsto h_{P_A} (P_A; \lambda_{P_A}),$ where each of polymerization functions can be chosen to be the identity function and no further restrictions apply to them.
\end{itemize}
\noindent For a system satisfying the assumptions 1-5 listed above gauge fixing and polymerization do not commute.
\end{customlemma}
\begin{customproof}{of lemma 2}
~\\
\normalfont
Without loss of generality let $G_0$ be the specified gauge fixing condition in the lemma. Then we can compute its temporal derivative in the effective theory
\begin{eqnarray}
    \frac{\dd}{\dd t}G_0(t,x) &=& \{G_0(t,x),\Hcaneff\} + \frac{\partial G_0}{\partial t}(t,x)\nb \\
   & =& \int d^3z^{\prime\prime}\int d^3z^\prime\int d^3z\int d^3y \Big(\frac{\partial h_{Q^0}(z^\prime)}{\partial Q^0(z)}\Big) \Big(\frac{\partial h_{P_0}(z^{\prime\prime})}{\partial P_0(z)}\Big) N^I(y) \nb\\ &&\qty(\fdv{G_0(t,x)}{h_{Q^0}(z^\prime)}\fdv{C_I(y)}{h_{P_0}(z^{\prime\prime})}- \fdv{G_0(t,x)}{h_{P_0}(z^{\prime\prime})}\fdv{C_I(y)}{h_{Q^0}(z^\prime)}) + \frac{\partial G_0(t,x)}{\partial t},
\end{eqnarray}
where we have used assumption 3 after the last line. Considering assumption 4 and 5 we have that  $\frac{\partial G_0}{\partial t}(t,x)$ as well as $\qty(\fdv{G_0(t,x)}{h_{Q^0}(z^\prime)}\fdv{C_I(y)}{h_{P_0}(z^{\prime\prime})}- \fdv{G_0(t,x)}{h_{P_0}(z^{\prime\prime})}\fdv{C_I(y)}{h_{Q^0}(z^\prime)})$ coincide with their polymerized classical counterparts. Since by assumption 1 the classical solutions for the Lagrange multipliers $N^I_{\rm class}$ exist, we can reinsert the polymerized classical solution denoted by $N^I_{\rm class}$(h) into the above result and obtain
\begin{eqnarray}
   && \int d^3z^{\prime\prime}\int d^3z^\prime\int d^3z\int d^3y \Big(\frac{\partial h_{Q^0}(z^\prime)}{\partial Q^0(z)}\Big) \Big(\frac{\partial h_{P_0}(z^{\prime\prime})}{\partial P_0(z)}\Big) N^I(y) \nb\\ &&\qty(\fdv{G_0(t,x)}{h_{Q^0}(z^\prime)}\fdv{C_I(y)}{h_{P_0}(z^{\prime\prime})}- \fdv{G_0(t,x)}{h_{P_0}(z^{\prime\prime})}\fdv{C_I(y)}{h_{Q^0}(z^\prime)}) + \frac{\partial G_0(t,x)}{\partial t},\nb\\
    &\approx&
\int d^3z^{\prime\prime}\int d^3z^\prime\Big[ \int d^3z\Big(\frac{\partial h_{Q^0}(z^\prime)}{\partial Q^0(z)}\Big) \Big(\frac{\partial h_{P_0}(z^{\prime\prime})}{\partial P_0(z)}\Big)-\delta^{(3)}(z^\prime,z^{\prime\prime})\Big] \int d^3yN^I_{\rm class}(h)(y) \nb\\ &&\qty(\fdv{G_0(t,x)}{h_{Q^0}(z^\prime)}\fdv{C_I(y)}{h_{P_0}(z^{\prime\prime})}- \fdv{G_0(t,x)}{h_{P_0}(z^{\prime\prime})}\fdv{C_I(y)}{h_{Q^0}(z^\prime)}).
\end{eqnarray}
This expression cannot weakly vanish because on the one hand using assumption 3 we have that at most one polymerization function can be the identity function and moreover we have $\int d^3z\Big(\frac{\partial h_{Q^0}(z^\prime)}{\partial Q^0(z)}\Big) \Big(\frac{\partial h_{P_0}(z^{\prime\prime})}{\partial P_0(z)}\Big)\not\approx \delta^{(3)}(z^\prime,z^{\prime\prime})$.  On the other hand due to assumption 1, the last integral on the RHS in the above equation  
cannot weakly vanish because then already in the classical theory the dynamical stability of the gauge fixing condition can not be satisfied as we have a non-vanishing $\frac{\partial G_I}{\partial t}\not\approx 0$ required in assumption 2.$\qedsymbol$
\end{customproof}
Note that we have considered the case of the field theory here. For models with a finite number of degrees of freedom the same result holds and the proof works similarly in this simpler setup. This is the reason why we have focused on field theory here.

\subsection{Reverse engineering gauge fixing conditions for given lapse and shift}
\label{sec:reverse engineering}
In this subsection we want to discuss the question whether one can take the point of view that for given choice of a polymerization of lapse and shift one can always reverse engineer a set of gauge-fixing conditions that are consistent with the effective dynamics and the polymerized lapse and shift. An application where this question becomes relevant is if elementary variables involved on the one side in the constraints and on the other side in lapse and shift are polymerized differently where both ways of polymerizing still have the correct classical limit. One of the main reasons why a different polymerization for the shift vector is  chosen is, that it yields a quantization where the shift vector operator acts on the same lattice points as the Hamiltonian constraint, see for eg. the discussion in \cite{Gambini:2020nsf}. A further independent argument for this choice is discussed in \cite{Kelly:2020uwj} where such a definition of the shift vector is favored to allow the algebra of the effective Hamiltonian constraints and the classical ones exactly agree. Although it is desirable to have no anomalies in this algebra, the first requirement that needs to be satisfied here is to choose a dynamically consistent shift vector. If there are ambiguities left these could be used to obtain an anomaly-free algebra but as we will see below for the model in \cite{Kelly:2020uwj,Gambini:2020nsf,Husain:2021ojz} a dynamically consistent shift vectors needs to have the same polymerization as the Hamiltonian and diffeomorphism constraint or the physical Hamiltonian respectively.
Because allowing to adopt the gauge fixing conditions to a given set of lapse and shift brings some additional freedom and thus ambiguity with it, it is rather difficult to provide a general proof that is based on very general assumptions. Therefore, what we will present in our discussion below is a lemma that demonstrates that for a certain class of models available in the literature \cite{Kelly:2020lec,Gambini:2020nsf, Husain:2021ojz} such kind of reverse engineering is not possible. As a consequence, this means that the polymerization chosen for the constraints and thus the physical Hamiltonian needs also to be chosen for lapse and shift if one wants to have a consistent dynamical system.
At the end of this subsection we will comment on possible drawbacks that can occur if one follows the route of first choosing a polymerization for lapse and shift and then aims at reverse engineering the corresponding consistent gauge fixing conditions.

The class of models that we want to consider satisfy the following assumptions:
\begin{itemize}
\item[(a)] The gauge fixing condition associated with the Hamiltonian constraint denoted by $G_0$ depends only on the matter degrees of freedom denoted by $\Phi^{A},\pi_A$ and we have $\frac{\partial}{\partial t} G_0(\Phi^{A},\pi_A,t)\not\approx 0$.
\item[(b)] The matter degrees of freedom $\Phi^{A},\pi_A$  are not polymerized at the level of the effective dynamics.
\item[(c)] The gauge fixing conditions associated with the spatial diffeomorphism constraint $G_a$ do only depend on the gravitational triad variables and $\frac{\partial}{\partial t} G_a(E^a_j)\approx 0$.
\item[(d)] The polymerization of the connection variables denoted by $A^j_a$ in the constraints is performed by $A^j_a\mapsto h(A ^a_j;\lambda)$ where $h(A ^a_j;\lambda)$ does not depend on partial derivatives of $A^j_a$ and is not the identity function. The triad variables $E^a_j$ are not polymerized at the effective level.
\end{itemize}

The above assumptions  are for instance satisfied in the models considered in \cite{Kelly:2020lec,Husain:2021ojz}.  If we wish to reverse engineer the gauge fixing conditions for a given lapse and shift then in general this will also include the freedom to add extra polymerized contribution in the gauge fixing condition that vanish in the classical limit or multiply quantities in the gauge fixing condition with polymerized quantities that tend to one in the classical limit. With the chosen conditions for the polymerization in (d) we want to exclude these ambiguities here. Below we will discuss the situation when these assumptions are relaxed and particularly emphasize what kind of drawbacks can occur if one follows this route.
Given this set of assumptions we can prove the following lemma:
\begin{customlemma}{3}
\label{lemma4}
~\\
Given the assumptions (a)-(d) listed above, then the gravitational degrees involved in lapse and shift need to be polymerized in exactly the same way as chosen for the constraints and the physical Hamiltonian respectively. Modifying  the gauge fixing condition to allow different polymerization for lapse and/or shift is not possible if we require the effective system to be dynamically consistent and to possess the correct classical limit.
\end{customlemma}

\begin{customproof}{of lemma 3}
~\\
\normalfont
First we realize that the assumptions (a)-(d) are a special case of the assumptions stated in Corollary \ref{lemma3}. Hence, by applying Corollary \ref{lemma3} we can conclude that under these assumptions gauge fixing and polymerization commute. Next, by assumption (a) and (c) we know that all gauge fixing conditions are not affected by the polymerization and we have at the effective level
\begin{equation}
G_0 = G_0^{\rm class}(\Phi^{A},\pi_A;t)\quad{\rm and}\quad G_a=G_a^{\rm class}(E^a_j).
\end{equation}
We use the same notation as in the former proofs, then the stability condition reads
\begin{eqnarray}
\int d^3y \int d^3z \Neff^J \Feff_{\Neff^J,I}(x,y,z) +\delta_{I,0}\frac{\partial G_0}{\partial t}(t,x)&\approx & 0 ,\quad I,J=0,\cdots 3.
\end{eqnarray}
where we have used assumption (c) so far and considered further the assumptions (a),(b) and (d). The explicit form of $\Feff_{\Neff^J,I}(x,y,z)$  for this class of models is given by
\begin{eqnarray}
\label{eq:FormofF}
\Feff_{\Neff^J,0}(x,y,z)&=&   \FCl_{\Neff^J,0}[\Phi^A, \pi_A](x,y,z)\quad{\rm and} \nonumber\\
\Feff_{\Neff^J,a}(x,y,z) &=&  \FCl_{\Neff^J,a}[\polyF,E^a_j,\Phi^A, \pi_A](x,y,z){\cal H}(A^j_a),
\end{eqnarray}
with
\begin{eqnarray}
\FCl_{\Neff^J,0}[\Phi^A, \pi_A](x,y,z)&=&  \frac{\delta G_0(t,x)}{\delta \Phi^A(z)}\frac{\delta \Ccl_{J,{\rm mat}}(y)}{\delta \pi_A(z)} - \frac{\delta G_0(t,x)}{\delta \pi_A(z)}\frac{\delta \Ccl_{J,{\rm mat}}(y)}{\delta \Phi^A(z)},\\
\FCl_{\Neff^J,a}[\polyF,E^a_j,\Phi^A, \pi_A](x,y,z)&=&
- \frac{\delta G_a(t,x)}{\delta E_k^b(z)}\frac{\delta \Ccl_{J,{\rm geo}}(y)}{\delta A_b^k(z)}\Big|_{A^j_a\to \polyF},
\end{eqnarray}
where $\Ccl_{J,{\rm mat}}$ and $\Ccl_{J,{\rm geo}}$ denote the matter and geometrical contribution to $\Ccl_J$ and we further used that gauge fixing and polymerization commute for this class of models and computed the analogue of \eqref{eq:CondCommute} in Lemma \ref{lemma} for the models considered here. ${\cal H}(A^j_a)$ is as before a generic function that depends on the connection variables and its explicit form will depend on the chosen polymerization. Using \eqref{eq:FormofF} we obtain the following equation that determines the effective lapse $N$
\begin{equation}
\int\limits d^3y \int\limits d^3z N\FCl_{\Neff ,0}(\Phi^{A},\pi_A)(x,y,z)
\approx-\int\limits d^3y \int\limits d^3z N^a\FCl_{\Naeff{a} ,0}(\Phi^{A},\pi_A)(x,y,z)-\delta_{I,0}\frac{\partial G_0}{\partial t}(t,x).\\
\end{equation}
This agrees exactly with the equation we obtain at the classical level up to the fact that instead of the classical shift vector $\Nacl{a}$, the effective one $\Naeff{a}$ is involved. Therefore, we can conclude that for the models considered here, we always have the lapse  function of the effective theory  given by $\Neff=\Ncl(\Phi^{A},\pi_A,\Naeff{a})$ and polymerized variables  involved in $\Neff$ can only come from $\Naeff{a}$. Substituting this into the three remaining equations that determine the effective lapse function this yields for $a=1,2,3$
\begin{flalign}
\int\limits d^3y \int\limits d^3z \Naeff{b}\FCl_{\Naeff{b} ,a}(\polyF,E^a_j,\Phi^{A},\pi_A)(x,y,z)
&\approx& \nb\\
&& \hskip-5cm-\int\limits d^3y \int\limits d^3z \Ncl(q^A,P_A,\Naeff{a})\FCl_{\Neff ,a}(\polyF,E^a_j,\Phi^{A},\pi_A)(x,y,z) .
\end{flalign}
Since by assumptions (b) and (c) the gauge fixing conditions do not involve any polymerized variables, it follows that the polymerization encoded in $\FCl_{\Naeff{b} ,a}$ and $\FCl_{\Neff ,a}$ is completely determined by the kind of polymerization chosen for the Hamiltonian and spatial diffeomorphism constraints. Moreover, the polymerized variables involved in the lapse function can only come from the contribution of $\Naeff{a}$ and therefore any involved polymerized variables in the system of equations that determine the effective lapse and shift are contributions from the polymerized variables involved in the effective constraints.    Since the  effective physical Hamiltonian in these kind of models will always be a phase space function that involves contributions of the effective constraints from the gravitational degrees of freedom and some matter degrees of freedom, the latter contribution depending on the explicit form of the chosen $G_0$, the polymerization chosen for the gravitational part of the constraints carries over to the effective physical Hamiltonian as well. Now due to the weak equality in the system of equations that determine lapse and shift at the effective level, we can use the constraints at the effective level to for instance replace certain matter variables by geometrical ones that are polymerized at the effective level. But if this is done, the polymerization involved in the final result for lapse and shift is again completely determined by the polymerization chosen for the constraints. This  shows that starting with a given gauge fixing condition that is consistent with the assumptions (a)-(d) the polymerization of lapse and shift is determined by the polymerization chosen for the constraints and cannot be chosen independently of this choice if we require dynamical consistency. Moreover, in these models we do also not have the freedom to modify the gauge fixing condition consistently to allow a different polymerization for lapse and shift than chosen for the constraints for the following reason: by assumption (b) and (d) neither the triads nor the matter variables experience any polymerization. For this reason for models satisfying assumption (a)-(d) it is not possible to modify the set of gauge-fixing conditions in a way that they on the one hand lead to a solution of the stability requirement that allows a different polymerization for lapse and/or shift -and are thus dynamically consistent- and on the other hand still have the correct classical limit. $\qedsymbol$
\end{customproof}

To close this subsection we want to look at a more general case. This means we will drop most of the assumptions of Lemma \ref{lemma4} and consider an effective system where the components of the connection in the constraints were polymerized according to $A^j_a\mapsto\polyF$. The triad variables $E^a_j$ as well as the matter degrees of freedom $\Phi^{A},\pi_A$ are not polymerized. Given a choice of lapse and shift we want to reverse engineer the corresponding gauge fixing conditions $G_I(A^j_a, E^b_i,\Phi^{A}, \pi_A; t, x, \lambda^j_a)$. In order to do that we have to consider the dynamical stability of the gauge fixing conditions with respect to the canonical Hamiltonian. Using the notation of foregone proofs this system of equations takes the form
\begin{align}
   \int d^3y \int d^3z N^J(y)\qty(\frac{\delta G_I(t,x)}{Q^\alpha(z)}\frac{\delta \Caeff{J} (y)}{\delta P_\alpha(z)} - \frac{\delta G_I(t,x)}{\delta P_\alpha(z)}\frac{\delta \Caeff{J}(y)}{\delta Q^\alpha(z)}) +\frac{\partial G_I}{\partial t}(t,x)\approx 0,
\end{align}
where $Q^\alpha=(A^j_a,\Phi^A)$ denotes all configuration variables of the phase space and $P_\alpha=(E^a_j,\pi_A)$ the respective canonically conjugate momenta. Additional factors of the Poisson algebra were in the above equation absorbed in the constraints. We can see that in general this is a very complicated system of equations. Further we only have a weak equality, meaning we are allowed to set the constraints on the left hand side to zero. A common strategy to find solutions is to make an ansatz for the gauge fixing conditions. Typically we want to restrict the dependence on the canonical variables or on the temporal variable. We use such a strategy in the first example of the following Sec. \ref{sec:Schwarzschild interior}. Note that we can already see in this example that in general there is no unique solution for the gauge fixing conditions.

Assuming we have found a solution to the above system of equations $G_I^{\rm sol}$, then we want to formulate in the following various sets of criteria the solution has to meet and discuss their implications.
First we require the solution for the gauge fixing conditions to have the correct classical limit and obey the polymerization that was already used in the constraints (in the following we refer to such a polymerization as standard). In our case the latter criterion means that there exists a function $\widetilde{G_I}\qty(Q^\alpha, P_\alpha;t,x)$ such that
\begin{equation}
    \widetilde{G_I}(\polyF, E^b_i,\Phi^{A}, \pi_A;t,x)= G_I^{\rm sol}(A^j_a, E^b_i,\Phi^{A}, \pi_A; t, x, \lambda^j_a).
\end{equation}
This is the simplest case, since we have a consistent system where the effective physical Hamiltonian is equal to the standard polymerization of the classical physical Hamiltonian. Practically we can directly check whether such a solution exists for a given system by plugging in the standard polymerized classical gauge fixing conditions as well as our choice for lapse and shift in the stability equations and check if they weakly vanish.

Secondly we can relax the criterion for the standard polymerization and solely demand the solution for gauge fixing conditions to have the correct classical limit. Note that the model discussed in Sec. \ref{sec:Schwarzschild interior} falls into this category. For this class of solutions we have to be careful when formulating the effective physical Hamiltonian. Instead of using the standard polymerized gauge fixing conditions we have to use our solution and the choice of lapse and shift to construct the physical Hamiltonian from the polymerized canonical Hamiltonian $\Hcaneff$. Note that although the classical limit of such a physical Hamiltonian is correct, it is not given by the standard polymerization of the classical physical Hamiltonian. The price to pay for this approach is that we need to allow any possible quantization ambiguity in the gauge fixing conditions to obtain consistent solutions. Since such an ambiguity is considered for the quantization of the gauge fixing condition only, the question remains why the gauge fixing conditions are quantized differently from all other functions.

The third category of solutions is defined by having a classical limit that does not agree with the classical gauge fixing condition of the original classical model. As a consequence, the physical Hamiltonian will also have a different classical limit. This means that these effective gauge fixing conditions correspond to a different classical system than the one we originally wanted to construct an effective theory thereof. In a sense different choices of gauge fixing conditions do not change fundamentally the physics of the theory but rather the perspective from which we describe it, for example a rescaling of the temporal coordinate. Still, if we want to have a system for which the effective theory is a direct extension of the classical one, we need to replace the classical theory by the system constructed from the classical limit of the gauge fixing conditions and take into account that the effective theory corresponds to a classical theory with a different chosen gauge condition than we originally started with at the classical level. Such a perspective could be followed if one takes the quantum theory and its corresponding effective model as fundamental and not as the quantization of a given classical model because the latter is problematic if we cannot rediscover such classical theory in the classical limit. Depending on whether the polymerization of the gauge fixing conditions is standard or not we would obtain models which fits in the first or second category.

As our results show, from a physical perspective if we choose lapse and shift and then reverse engineer consistent gauge fixing conditions a wide range of the choices for lapse and shift are suboptimal in the following sense. In models where polymerization and gauge fixing do not commute, in general in order to work with a consistent model we need to rely on quantization ambiguities (second category) and work with polymerized gauge fixing conditions whose classical limit does not agree with the classical gauge fixing condition one originally has started with  (third category).

\section{Examples from loop quantum black holes and spherically symmetric spacetimes}
\label{sec:Examples}
To illustrate the consistency requirements of gauge fixing conditions with respect to the polymerized gravitational dynamics we consider three specific examples. We first discuss the case of the loop quantization of the Schwarzschild interior using the quantization proposed in \cite{Corichi:2015xia}.\footnote{While there exist more recent loop quantizations of the Kruskal spacetime such as \cite{Ashtekar:2018lag, Ashtekar:2018cay} which not only resolve the central singularity but are also free from some undesirable features such as large mass asymmetry between black hole and white hole spacetimes across the bounce, we consider this quantization due to its simplicity and some features which are shared by others \cite{Olmedo:2017lvt,Ashtekar:2018lag, Ashtekar:2018cay, Bodendorfer:2019cyv}. } In this quantization all of the assumptions for Lemma \ref{lemma} in the previous section are satisfied. We find that polymerization and  gauge fixing  do not  commute. Then, we move on to our second example where the gravitational collapse of an inhomogeneous dust cloud in the LTB spacetime is considered. In this example, which satisfies the assumptions of Corollary \ref{lemma3}, one works with two gauge fixing conditions, one of them being a temporal gauge fixed by a reference field. We illustrate the conditions under which gauge fixing and polymerization commute and point out issues with considering different polymerization for the shift vector as considered in \cite{Kelly:2020lec} which is not consistent with stability of gauge fixing conditions. In our third example we discuss the case of matter clocks when polymerization of matter fields is taken into account. Lemma \ref{corollary3} shows in this case that polymerization and gauge fixing do not commute.

\subsection{The Schwarzschild interior}
\label{sec:Schwarzschild interior}

Our first example concerns the interior of the Schwarzschild black hole which is isometric to the homogeneous and anisotropic Kantowski-Sachs spacetime. Due to the underlying symmetries this can be understood as a special case of our proof where we consider a system with finitely many degrees of freedom and where the diffeomorphism constraint is trivially satisfied such that one deals with only one gauge fixing condition.  In this model the spacetime metric is given by
\bq
\label{KS spacetime}
\dd s^2=-N_\mathrm{class}(t)^2\dd t^2+f(t)\dd^2x+g(t)\dd^2\Omega,
\eq
here $N_\mathrm{class}$ denotes the lapse function, $x$ is the radial coordinate and $d^2\Omega=d\theta^2+\sin^2(\theta)d\phi^2$ represents the angular part of the metric. For the  Schwarzschild interior,
\bq
N_\mathrm{class}=\frac{1}{2m/t-1}, \quad f=N^2(t),\quad g(t)=t^2,
\eq
where $m=G M$ and $M$ is the mass of the black hole. The coordinates can take any values in the range $t\in[0,2m), x\in \mathbb{R},\theta\in[0,\pi],\phi\in[0,2\pi]$.

In the following we first discuss the case of the classical theory which is followed by its polymerized version.

\subsubsection{The gauge fixing conditions in the classical Schwarzschild interior}

It can be shown that in the Kantowski-Sachs spacetime, after imposing the Gauss constraint, the Ashtekar connection and the densitized triads only depends on two pairs of symmetry reduced canonical variables, namely, $(b,p_b)$ and $(c,p_c)$ \cite{Ashtekar:2005qt}. These phase space variables satisfy the following non-vanishing Poisson brackets:
\bq
\{b,p_b\}=G\gamma, \quad \quad \{c,p_c\}=2G\gamma.
\eq
The corresponding classical Hamiltonian for the  Schwarzschild interior in terms of canonical pairs $(b,p_b)$ and $(c,p_c)$ takes the form
\bq
\label{interior ham}
\Hcancl=-\frac{\Ncl}{2G\gamma^2}\Big[\left(b^2+\gamma^2\right)\frac{p_b}{\sqrt p_c}+2bc\sqrt p_c\Big] .
\eq
Interestingly, the lapse $\Ncl$ can be chosen in such a way that the resulting Hamilton's equations can be readily solved and the physical interpretations of the solutions are transparent \cite{Ashtekar:2005qt,Corichi:2015xia}. In particular, when
\bq
\label{lapse1}
\Ncl=\gamma \sqrt p_c/b,
\eq
the equations of motion of $(b,p_b)$ decouple from those of $(c,p_c)$\footnote{It can be shown that multiplying the classical lapse (\ref{lapse1}) by a function of $b$ or $c$ would lead to the Hamilton's equations which still admit analytical solutions, the additional function which is multiplied to (\ref{lapse1}) amounts to a redefinition of time.}, it is then straightforward to obtain the general solutions \cite{Corichi:2015xia}
\bqn
\label{solutions1}
p_b&=&p^{(0)}_be^t\sqrt{e^{-t}-1},\quad p_c=p^{(0)}_c e^{2t},\nb\\
b&=&\pm\gamma\sqrt{e^{-t}-1},\quad \quad c=c_0e^{-2t},
\eqn
where the coordinate $t$ should be regarded as tailored to the lapse (\ref{lapse1}). Since the diffeomorphism constraint is already fixed in obtaining this model, a generic form of the gauge fixing condition which leads to above particular lapse function  must explicitly depend on the time coordinate. We consider the following ansatz for the gauge fixing condition:
\bq
\label{GF}
G^\mathrm{class}_0=f_0(c,p_c,b,p_b)-t .
\eq
Here $f_0$ is a function of the canonical variables which does not depend explicitly on the coordinate time. By requiring the preservation of the gauge fixing condition (\ref{GF}) with respect to the Hamiltonian (\ref{interior ham}), namely
\bq
\frac{\mathrm{d}}{{\mathrm{d} t}} G^\mathrm{class}_0 = \{ f_0, \Hcancl \} - 1 \approx 0,
\eq
one can readily obtain a partial differential equation (PDE) for $f_0$. For the lapse in (\ref{lapse1}), one gets:
\bq
\label{PDE1}
-\left(\frac{b^2+\gamma^2}{2b}\right)\frac{\partial f_0}{\partial b}+\left(p_b-\frac{b^2+\gamma^2}{2b}\right)\frac{\partial f_0}{\partial p_b}-2c\frac{\partial f_0}{\partial c}+2p_c\frac{\partial f_0}{\partial p_c}=1.
\eq
Note that every solution of the above PDE  results in the same lapse function given in (\ref{lapse1}). Thus, there does not exist a one to one correspondence between the lapse and the gauge fixing conditions, and, in general, there can be different choices of the gauge fixing conditions which correspond to a particular lapse.

Some simple solutions of  (\ref{PDE1}) can be found easily. For example, when $f_0$ only depends on one of the canonical variables $b$, $c$, $p_c$ we get the following cases.\footnote{Given the PDE (\ref{PDE1}) there is no case when $f_0$ depends only on $p_b$.} \\

{\bf Case A:} When $f_0$ only depends on $c$, one can find that the corresponding solution of PDE (\ref{PDE1}) leads to the gauge fixing condition
\bq
\label{gf1}
G^\mathrm{A,class}_0=-\frac{1}{2}\ln\left(\frac{c}{c_0}\right)-t= 0.
\eq
As a result, when imposing the gauge fixing condition $G^\mathrm{A,class}_0=0$, we recover $c=c_0e^{-2t}$ which is one of the solutions in (\ref{solutions1}) derived from the Hamilton's equations for the lapse (\ref{lapse1}). \\

{\bf Case B:} When $f_0$ depends only on $p_c$, we find the corresponding solution results in the gauge fixing condition
\bq
\label{gf12}
G^\mathrm{B,class}_0=-\frac{1}{2}\ln\left(\frac{p_c}{p^{0}_c}\right)-t=0.
\eq
In this case, imposing the gauge fixing condition $G^\mathrm{B,class}_0=0$ leads to the solution for $p_c$, namely $p_c=p^{0}_c e^{2t}$  which is also given in (\ref{solutions1}).\\

{\bf Case C:} When $f_0$ only depends on $b$, solving the PDE (\ref{PDE1}) of $f_0$ leads to the gauge fixing condition
\bq
\label{gf2}
G^\mathrm{C,class}_0=-\ln \left(\frac{b^2+\gamma^2}{\gamma^2}\right)-t=0.
\eq
In this case, imposing the gauge fixing condition $G^\mathrm{C,class}_0=0$ leads to the solution for $b$, namely  $b=\pm\gamma\sqrt{e^{-t}-1}$ in (\ref{solutions1}).\\

Note that we have adjusted the integration constant in each case appropriately so that the corresponding gauge fixing conditions are exactly equivalent to the classical solutions in (\ref{solutions1}). In principle, any two gauge fixing conditions which differ by a constant will lead to the same lapse function. Hence, without any loss of generality, one can always properly choose this integration constant to make the form of the gauge fixing conditions consistent with the analytical solutions.

\subsubsection{The gauge fixing conditions in the polymerized Schwarzschild interior}
\label{sec:gaugeconditions_Schwarzschild}

The loop quantization of the Schwarzschild interior has been extensively studied in the literature \cite{Ashtekar:2005qt,Modesto:2005zm,Boehmer:2007ket,Corichi:2015xia,Olmedo:2017lvt,Ashtekar:2018lag,Ashtekar:2018cay} (see also \cite{Dadhich:2015ora} for Schwarzschild-de Sitter, Schwarzschild-anti-deSitter and higher genus black holes). In the following, we consider the polymerization in \cite{Corichi:2015xia} as an example to show that the gauge fixing conditions and the polymerizations do not commute according to Lemma \ref{lemma} and \ref{corollary3}. Following \cite{Corichi:2015xia}, the effective Hamiltonian of the loop quantized Schwarzschild interior is given by
\bq
\label{csham}
\Hcaneff=-\frac{N}{2 G\gamma^2}\Bigg[\left(\frac{\sin^2\left(\delta_b b\right)}{\delta^2_b}+\gamma^2\right)\frac{p_b}{\sqrt p_c}+2\frac{\sqrt p_c\sin\left(\delta_b b\right)\sin\left(\delta_c c\right)}{\delta_b\delta_c}\Bigg],
\eq
where two polymerization factors $\delta_b$ and $\delta_c$ are constants  which do not depend on the phase space variables. When these two polymerization factors approach zero, the effective Hamiltonian (\ref{csham}) tends to its classical counterpart in (\ref{interior ham}). Similar to the classical case discussed in the last subsection, one can then choose a particular lapse

\bq
\label{lapse2}
N=N^{\mathrm{poly}}=\frac{\gamma \delta_b \sqrt p_c}{\sin\left(\delta_b b\right)},
\eq
which decouple the $(b,p_b)$ sector from the  $(c,p_c)$ at the level of effective dynamics. Note that the lapse (\ref{lapse2}) is exactly the polymerization of the classical lapse in (\ref{lapse1}). The corresponding Hamilton's equations admit analytical solutions which are given explicitly by \cite{Corichi:2015xia}
\bqn
\label{analytical solutions}
c&=&\frac{2}{\delta_c}\arctan\left(\frac{\delta_c}{2}c_0e^{-2t}\right),\quad \quad b=\pm\frac{1}{\delta_b}\arccos\Bigg[b_0\tanh\left(\frac{1}{2}b_0t+\tanh^{-1}(1/b_0)\right)\Bigg],\nb\\
p_c&=&4m^2\left(e^{2t}+c^2_0\delta^2_c e^{-2t}\right),\quad\quad p_b=-2\frac{\sin(\delta_c c)}{\delta_c}\frac{\sin(\delta_b b)}{\delta_b}\frac{p_c}{\frac{\sin^2(\delta_b b)}{\delta^2_b}+\gamma^2},
\eqn
here $c_0$ is an integration constant and $b_0=\sqrt{1+\gamma^2\delta^2_b}$. When $\delta_b$ and $\delta_c$ approach zero, these solutions from the effective dynamics tend to their classical limits given in (\ref{solutions1}).

In order to obtain the gauge fixing conditions corresponding to the polymerized lapse (\ref{lapse2}), one can follow the procedure in the classical case where we assume that a general gauge fixing condition takes the form (\ref{GF}) and then demand it be preserved during the time evolution of the effective dynamics generated by the Hamiltonian  (\ref{csham}) with the lapse (\ref{lapse2}). In doing so, we obtain another PDE for $f_0$ which turns out to be
\bq
\label{PDE2}
-\frac{1}{2}\left(\frac{\sin(\delta_b b)}{\delta b}+\frac{\gamma^2 \delta_b}{\sin(\delta_b b)}\right)\frac{\partial f_0}{\partial b}+\alpha\frac{\partial f_0}{\partial p_b} -2\frac{\sin(\delta_c c)}{\delta_c}\frac{\partial f_0}{\partial c}+2p_c \cos(\delta_c c)\frac{\partial f_0}{\partial p_c}=1,
\eq
with $\alpha=\left(p_b\cos\left(\delta_b b\right)+\frac{\delta_b p_c \cos(\delta_b b)\sin(\delta_c c)}{\sin(\delta_c c)\delta_c}\right)$. Any gauge fixing conditions in the analogous form of (\ref{GF}) which are compatible with the effective Hamiltonian (\ref{csham}) together with the polymerized lapse (\ref{lapse2}) must satisfy the above PDE. We can obtain following particular solutions for the gauge fixing conditions from the above PDE for cases when $f_0$ depends only on $b$, $c$ and $p_c$.\\

{\bf Case A:} When $f_0$ only depends on the connection $c$, the solution to (\ref{PDE2}) results in the gauge fixing condition
\bq
\label{gf4}
G^A_0=-\frac{1}{2}\ln\left(\frac{\tan\left(\frac{\delta_c c}{2}\right)}{\frac{\delta_c c_0}{2}}\right)-t=0,
\eq
which reduces to (\ref{gf1}) as $\delta_c$ tends to zero on one hand and gives the analytical solution of $c$ in (\ref{analytical solutions}) on the other hand. Therefore, this gauge fixing condition is compatible with the dynamics when the lapse is chosen to be the one from the polymerization of the classical lapse. It is important to note that in the Hamiltonian constraint (\ref{csham}) and the lapse (\ref{lapse2}), the sine function is used to polymerize the connections while in the above gauge fixing condition, a tangent function appears as a polymerization of the connection $c$. The gauge fixing condition does not have the same polymerization as the one used in the lapse and the Hamiltonian constraint! This example illustrates what can happen in a general case. Due to Lemma \ref{lemma} we know that gauge fixing and polymerization do not commute. Solving above PDE we end up with a solution of the second category according to the discussion at the end of Sec. \ref{sec:reverse engineering}.\\

{\bf Case B:} When $f_0$  depends on $p_c$, the PDE (\ref{PDE2}) implies $f_0$ should depend on $c$ as well. In this case, we need to look for a solution of $f_0$ which depends on both $c$ and $p_c$. Such a solution exists and the corresponding gauge fixing condition is  given by
\bq
\label{gfb}
G^B_0=-\frac{1}{2}\ln\left( \frac{2\tan\left(\frac{\delta_c c}{2}\right)}{\delta_c}\right)+g\Big(p_c\frac{\sin(\delta_cc)}{\delta_c}\Big)-t=0,
\eq
where $g\Big(p_c\frac{\sin(\delta_cc)}{\delta_c}\Big)$ is an arbitrary differentiable function of $p_c\frac{\sin(\delta_cc)}{\delta_c}$. Given the solutions of $c$ and $p_c$ in (\ref{analytical solutions}), it is straightforward to show that $p_c\frac{\sin(\delta_cc)}{\delta_c}$ turns out to be a constant. As a result, the gauge fixing condition $G^B_0=0$ is also a solution of the effective dynamics with the lapse (\ref{lapse2}). Note the gauge fixing condition (\ref{gfb}) for the effective dynamics does not correspond to the polymerization of its classical counterpart (\ref{gf12}) in the sense that the latter only depends on the momentum variable $p_c$ while the combination $p_c\frac{\sin(\delta_cc)}{\delta_c}$  whose classical counterpart is $cp_c$ always show up together in $G^B_0$. This solution falls into the third category since the classical limit is different from the classical gauge fixing condition. \\

{\bf Case C:} When $f_0$ only depends on the triad variable $b$, we can keep  the first term on the left-hand side of the PDE (\ref{PDE2}) to obtain a solution of $f_0$, which leads to the gauge fixing condition
\bq
\label{gf5}
G^C_0=\frac{2}{b}\tanh^{-1}\left(\cos(\delta_b b)/b_0\right)-\frac{2}{b_0}\tanh^{-1}(1/b_0)-t= 0,
\eq
where $b_0=\sqrt{1+\gamma^2\delta^2_b}$ and the above gauge fixing condition exactly gives the analytical solution of $b$ in (\ref{analytical solutions}). Moreover, when $\delta_b$ approaches zero, the gauge fixing condition (\ref{gf5}) recovers its classical limit given in (\ref{gf2}). As can be clearly seen, $G^C_0$ can not be obtained from the polymerization of the connection $b$ in the classical gauge fixing condition (\ref{gf2}). Instead it can only be obtained by requiring its compatibility with the effective dynamics.

To summarize these examples, we have explicitly shown that our Lemma \ref{lemma} applies to the Schwarzschild interior where the polymerization of the lapse and the Hamiltonian constraint  lead to  consistent gauge fixing conditions which do not directly come from the polymerizations of their classical counterparts but are obtained by demanding the dynamical consistency of the gauge fixing conditions. Requiring this consistency of the effective dynamics shows that polymerization does not commute with the gauge fixing.

A pertinent question is the following. At the level of the effective dynamics, what are the lapse functions which are consistent with the gauge fixing conditions that come directly from the polymerization of the classical ones in (\ref{gf1})-(\ref{gf2})? For the cases discussed above, these polymerized gauge fixing conditions obtained directly from the classical gauge fixing conditions, take the forms,
\bq
\label{gf3}
G^\mathrm{A,poly}_0=-\frac{1}{2}\ln\left(\frac{\sin\left(\delta_c c\right)}{c_0 \delta_c}\right)-t,\quad  G^\mathrm{B,poly}_0=-\frac{1}{2}\ln\left(\frac{p_c}{p^{(0)}_c}\right)-t,\quad  G^\mathrm{C,poly}_0=-\ln \left(1+\frac{\sin^2(\delta_b b)}{\delta^2_b\gamma^2}\right)-t.
\eq
Requiring the preservation of the above polymerized gauge fixing conditions under the effective dynamics produced by the effective Hamiltonian (\ref{csham}), we can readily obtain a PDE of the lapse for each gauge fixing condition in (\ref{gf3}). Since the calculations for each gauge fixing condition is similar, in the following, we take $G^\mathrm{A,poly}_0$ as an example. Consistency requirement demands that
\bq
\label{fixation of the effective lapse}
\frac{\mathrm{d}}{\mathrm{d} t}  G^\mathrm{A,poly}_0=\{G^\mathrm{A,poly}_0, \Hcaneff\}+\frac{\partial}{\partial t} G^\mathrm{A,poly}_0=N\frac{\sin(\delta_b b)\cos{\delta_c c}}{\gamma \delta_b \sqrt p_c}-1=0,
\eq
which immediately yields
\bq
\label{lapse3}
N=\frac{\gamma \delta_b \sqrt p_c}{\sin(\delta_b b)\cos(\delta_c c)}.
\eq
It is straightforward to show that for the gauge fixing conditions given in (\ref{gf3}), the consistent lapse function are all given by (\ref{lapse3}).
Although this lapse also tends to the  classical lapse (\ref{lapse1}) when $\delta_b$ and $\delta_c$ approach zero, due to the additional cosine function in the denominator, it is not the polymerization of the classical lapse under the same polymerization rules as used in the Hamiltonian constraint and the gauge fixing conditions. In this sense, using the same polymerizations in the classical lapse and the classical gauge fixing conditions are not compatible with the effective dynamics governed by the effective Hamiltonian (\ref{csham}). Last but not the least, we can directly find the non-commutativity between the gauge fixing and the lapse  by comparing the algebraic equations for the classical lapse and its counterpart in the effective theory. In the classical theory, knowing the gauge fixing condition $G^\mathrm{A,class}_0=0$, the classical lapse is determined by
\bq
\label{fixation of the classical lapse}
\frac{\mathrm{d}}{\mathrm{d} t} G^\mathrm{A,class}_0=\{G^\mathrm{A,class}_0, \Hcaneff\}+\frac{\partial}{\partial t} G^\mathrm{A,class}_0=\frac{Nb}{\gamma \sqrt{p_c}}-1=0.
\eq
As we can see now clearly that the polymerization of equation (\ref{fixation of the classical lapse}) would not lead to its counterpart which is (\ref{fixation of the effective lapse}) in the effective theory due to the additional cosine function in the latter which is exactly the factor ${\cal H}(A^c_\ell)$ in (\ref{polymerizationsimple}) introduced in the proof of Lemma \ref{lemma}.

\subsection{Gravitational collapse of an inhomogeneous dust cloud in the LTB spacetime}
\label{JorgeEd_model}
Spherically symmetric spacetimes were first quantized using complex Ashtekar variables in \cite{Thiemann:1992jj} and real Ashtekar-Barbero variables in \cite{Bojowald:2005cb}. These techniques have been used in recent years to explore quantum geometric effects in various spherically symmetric spacetimes \cite{Campiglia:2007pb,Gambini:2008dy,Tibrewala:2012xb,Chiou:2012pg, Han:2020uhb,Kelly:2020uwj, Benitez:2020szx, Gambini:2020nsf}. In this subsection, we 
 consider the gravitational collapse of an inhomogeneous dust cloud in a spherically symmetric spacetime.  This example fits into Corollary \ref{lemma3} of Lemma \ref{lemma2}, and Lemma \ref{corollary3}. The spacetime whose metric can be expressed in terms of the Ashtekar-Barbero variables as
\bq
ds^2=-N^2_\mathrm{class}dt^2+\frac{(E^b)^2}{E^a}\left(dx+N^x_\mathrm{class}dt\right)^2+E^ad\Omega^2,
\eq
where $N_\mathrm{class}$ and $N^x_\mathrm{class}$ are the lapse function and the radial component of the shift vector, $E^a(t,x)$ and $E^b(t,x)$ are the densitized triads in the radial and angular directions and $d\Omega^2=d\theta^2+\sin^2\theta d\phi^2$. In the following we first consider the classical case which is followed by its polymerized version.
After integrating out the angular part, the Hamiltonian of a collapsing dust cloud can be written in terms of Ashtekar-Barbero variables as \cite{Kelly:2020lec}
\begin{align}
\Hcancl = \int\mathrm{d}t\int \mathrm{d}x\, \Ncl \Ccl + \Nacl{x}\Cacl{x},
\end{align}
where the two remaining constraints are given by
\begin{align}
    \Ccl &= \tensor{C}{^{\text{geo}}} + \tensor{C}{^{\text{dust}}} = -\frac{1}{2 G \gamma^2}\qty(2 a b \sqrt{\tensor{E}{^a}} + \frac{\tensor{E}{^b}}{\sqrt{\tensor{E}{^a}}} (b^2 + \gamma^2)) + \frac{1}{8 G}\frac{\qty(\tensor{\partial}{_{x}}\tensor{E}{^a})^2}{\tensor{E}{^b}\sqrt{\tensor{E}{^a}}} + \frac{\sqrt{\tensor{E}{^a}}}{2G}\tensor{\partial}{_{x}} \qty(\frac{\tensor{\partial}{_{x}} \tensor{E}{^{a}}}{\tensor{E}{^b}})\nonumber \\
    & \quad\quad\quad\quad\quad\quad\quad\quad +4 \pi \tensor{p}{_T}\sqrt{1+\frac{\tensor{E}{^a}}{\qty(\tensor{E}{^b})^2}\qty(\tensor{\partial}{_x} T)^2},\\
    \Cacl{x} &= \tensor*{C}{^{\text{geo}}_x} + \tensor*{C}{^{\text{dust}}_x} = \frac{1}{2 G \gamma} \qty(2 \tensor{E}{^b} \tensor{\partial}{_x}b -   a\tensor{\partial}{_x}\tensor{E}{^a}) - 4 \pi \tensor{p}{_T}\tensor{\partial}{_x} T.
\end{align}
The Poisson algebra of the spherical symmetrized connection and triad as well as the dust field is given by \cite{Kelly:2020lec}
\begin{align}
    \poissonbracket{a(x)}{\tensor{E}{^a}(y)} &= 2 \gamma G \,\delta(x,y), \quad
    \poissonbracket{b(x)}{\tensor{E}{^b}(y)} &= \gamma G \,\delta(x,y), \quad
    \poissonbracket{T(x)}{\tensor{p}{_T}(y)} &= \frac{1}{4 \pi} \delta(x,y).
\end{align}
Since the system contains two first class constraints, namely, $\Ccl$ and $\Cacl{x}$, we need two gauge fixing conditions to convert the system into a second class system. The gauge fixing conditions we choose are
\bq
\label{gauge fixing conditions}
G_0=T - t,\quad \quad G_1= E^a - x^2.
\eq
The first of these gauge fixing conditions is also known as the dust-time gauge, which is typically used
for reference field models which incorporate matter in order to fix the gauge freedom and derive a physical Hamiltonian (see \cite{Giesel:2012rb} for a collection of such models). The second gauge condition $G_1=0$ is also referred to  as the areal gauge since it fixes radial coordinate $x$ to be the physical radius, i.e. a sphere of radius $x$ has the surface area $4 \pi x^2$.

The stability of the temporal gauge condition $\tensor{G}{_0}=0$ yields
\bq
\label{gauge1}
\dv{\tensor{G}{_0}}{t} = \poissonbracket{\tensor{G}{_0}}{\Hcancl} + \pdv{\tensor{G}{_0}}{t} = \Ncl \sqrt{1+\frac{\tensor{E}{^a}}{\qty(\tensor{E}{^b})^2}\qty(\tensor{\partial}{_x} T)^2} - \Nacl{x} \tensor{\partial}{_x} T -1 =0,
\eq
where we have used the dust-time gauge $T=t$ and thus $\tensor{\partial}{_x} T = 0$.  Considering the stability of the areal gauge, we find
\bq
\label{gauge2}
\dv{\tensor{G}{_1}}{t} = \poissonbracket{\tensor{G}{_1}}{\Hcancl} = 2 \Ncl\frac{b}{\gamma}  \sqrt{\tensor{E}{^a}} + \Nacl{x} \tensor{\partial}{_x}\tensor{E}{^a} =2 \Ncl  \frac{b}{\gamma} x + 2 \Nacl{x} x = 0 .
\eq
This gives a second equation which determines the lapse and shift. Solving the coupled equations (\ref{gauge1})-(\ref{gauge2}), one can obtain
\begin{align}
\Ncl= 1,\quad \quad  \Nacl{x} = - \frac{b}{\gamma}.
\end{align}
Having considered the classical case, we now want to analyze the stability of the gauge conditions in the effective theory. We can make this calculation for general polymerizations $A^j_a\mapsto \polyF$. Specifically, for connection variables $a$ and $b$, a general polymerization takes the form
\bq
\label{polymerization1}
a\mapsto h^a(a,\lambda_a)=\frac{\sin\left(\lambda_a a\right)}{\lambda_a } ,\quad \quad b\mapsto h^b(b,\lambda_b)=\frac{\sin\left(\lambda_b b\right)}{\lambda_b },
\eq
where $\lambda_a$ and $\lambda_b$ are polymerization factors which do not depend on the connection variables.
Correspondingly, the effective Hamiltonian now becomes
\bq
\Hcaneff = \int\mathrm{d}t\int \mathrm{d}x\, \Neff \Ceff + \Naeff{x}\Caeff{x},
\eq
with
\begin{align}
\Ceff&=-\frac{1}{2 G \gamma^2}\qty(2 h^a h^b \sqrt{\tensor{E}{^a}} + \frac{\tensor{E}{^b}}{\sqrt{\tensor{E}{^a}}} \left((h^b)^2 + \gamma^2\right)) + \frac{1}{8 G}\frac{\qty(\tensor{\partial}{_{x}}\tensor{E}{^a})^2}{\tensor{E}{^b}\sqrt{\tensor{E}{^a}}} + \frac{\sqrt{\tensor{E}{^a}}}{2G}\tensor{\partial}{_{x}} \qty(\frac{\tensor{\partial}{_{x}} \tensor{E}{^{a}}}{\tensor{E}{^b}})\nonumber \\
& \quad\quad\quad\quad\quad\quad\quad\quad +4 \pi \tensor{p}{_T}\sqrt{1+\frac{\tensor{E}{^a}}{\qty(\tensor{E}{^b})^2}\qty(\tensor{\partial}{_x} T)^2},\\
\Caeff{x} &=\frac{1}{2 G \gamma} \qty(2 \tensor{E}{^b} \tensor{\partial}{_x}h^b -h^a\tensor{\partial}{_x}\tensor{E}{^a}) - 4 \pi \tensor{p}{_T}\tensor{\partial}{_x} T.
\end{align}
Since the gauge fixing conditions (\ref{gauge fixing conditions}) does not depend on the connection variables, we continue to use them to fix the gauge freedom of the effective dynamics. Similar to the classical case, the preservation of the temporal and areal gauge conditions leads to a fixing of the lapse and shift. In particular, requiring the stability of the temporal gauge, we find
\bq
\dv{\tensor{G}{_0}}{t} = \poissonbracket{\tensor{G}{_0}}{\Hcaneff} + \pdv{\tensor{G}{_0}}{t}\approx N-1=0,
\eq
which is exactly the same as in the classical case since only the dust parts of the Hamiltonian have non-trivial contributions in the Poisson bracket. On the other hand,  the stability of the areal gauge requires
\bq
\dv{\tensor{G}{_1}}{t} = \poissonbracket{\tensor{G}{_1}}{\Hcaneff}= (\tensor{\partial}{_a}\tensor{h}{^{a}})\qty(2 \Neff\frac{\tensor{h}{^{b}}}{\gamma}  x + 2 \Naeff{x} x) =0,
\eq
where we have used the areal gauge after evaluating the Poisson bracket. The derivative of $h^a$ can be factored out as it only appears linearly in the Hamiltonian. Combining the last two  stability equations, one immediately finds
\begin{align}
\label{polymerized lapse and shift}
\Neff= 1,\quad \quad \quad  \Naeff{x} = - \frac{\tensor{h}{^{b}}}{\gamma}=-\frac{\sin\left(\lambda_b b\right)}{\lambda_b \gamma} .
\end{align}
The lapse function has the same value as in the classical theory and the shift vector is the polymerization of the classical shift vector. Note that in this model the temporal and the spatial gauges decouple. Since the stability of the temporal gauge condition is completely independent of the shift vector and the solution of the lapse function is the same in the classical as well as the effective theory, we can go to the partially gauge fixed system where only the spatial gauge remains. Then we can actually use Corollary \ref{lemma3} to show that gauge fixing and polymerization always commute. The important point is that now the system has no gauge condition with a temporal dependence, which is the exactly one of the conditions in Corollary \ref{lemma3}.

We conclude this subsection with two remarks on subtleties of using gauge fixing conditions which arise in recent works where this example is considered. These remarks help illustrate Lemma \ref{lemma4}. \\
\noindent
{\bf Remark 1:}
It should be noted that although the authors in \cite{Kelly:2020lec} use the same gauge fixing conditions given in (\ref{gauge fixing conditions}), they work with a different effective theory since the shift vector is polymerized differently than the Hamiltonian constraint. In particular, the shift vector used is (see also \cite{Kelly:2020uwj,Husain:2021ojz})
\bq
\label{shift2}
N^x=-\frac{1}{2\lambda_b \gamma} \sin\left(2\lambda_b b\right),
\eq
As compared with the shift vector in (\ref{polymerized lapse and shift}), the above shift has an additional factor of 2 in the argument of the trigonometric function. The motivation used to introduce such a polymerization is  to ensure that the operator of the scalar constraint as well as the shift vector act on the same vertices $\nu_j$. But such a choice is fraught with problems. First, it is rather unnatural and ad-hoc to choose a different polymerization exclusively for the shift vector. Secondly our above calculation shows that if we transition to the effective theory by polymerizing the canonical Hamiltonian and requiring stability of the polymerized gauge fixing conditions to determine the lapse and the shift of the effective theory, these solutions correspond to the polymerized classical solutions of the lapse and shift. One is \textit{not} allowed to choose any other polymerization as it will be inconsistent in the above sense. Thirdly, although the shift vector in (\ref{shift2}) has the same classical limit as the one in (\ref{polymerized lapse and shift}) when $\lambda_b$ tends to vanish, such a shift actually corresponds to a different choice of the gauge fixing conditions in the effective theory. Similar to what we have computed in the last subsection for the Schwarzschild interior,  the specific form of the gauge fixing condition consistent with the choice of the shift (\ref{shift2}) can be  obtained by requiring the preservation of the gauge fixing condition in the effective theory which leads to a partial differential equation of the gauge fixing condition. In summary we need to be careful in considering polymerizations of different quantities in the effective theory and choices made for certain simplifications can be inconsistent.  \\

\noindent
{\bf Remark 2:}
Another issue arises with respect to the choice of the lapse in (\ref{shift2}) when we consider the physical Hamiltonian of the model in \cite{Kelly:2020lec}. In principle, one can obtain the polymerized physical Hamiltonian in two different ways. The first is to impose the gauge fixing conditions (\ref{gauge fixing conditions}) in the classical theory  to get rid of one of the connection variables and its conjugate momentum and then use the polymerization in (\ref{polymerization1}) for the remaining connection variable. The other way is to first polymerize the full phase space with the ansatz for polymerization in (\ref{polymerization1}) and then impose the gauge fixing conditions (\ref{gauge fixing conditions}) to obtain the polymerized physical Hamiltonian. Since the gauge fixing and the polymerization commute in this model and the gauge fixing conditions remain the same for classical and effective theories, the polymerized physical Hamiltonian derived from the above two ways coincide. However, if one uses a different shift such as (\ref{shift2}) which corresponds to a different gauge fixing condition for the effective theory, then a consequence is that the above two ansatz would result in different polymerized physical Hamiltonians. In other words, in order to obtain  the polymerized physical Hamiltonian corresponding to the choice of the shift in (\ref{shift2}), one can only follow the second ansatz in which we first polymerize the full phase space then reduce to the physical phase space with the right gauge fixing conditions consistent with the shift (\ref{shift2}). This complexity does not arise if we simply choose the shift as given in (\ref{polymerized lapse and shift}) which is consistent with the choice of the gauge fixing conditions in (\ref{gauge fixing conditions}) in both classical and effective theories.

\subsection{Matter reference models including polymerized matter}
\label{sec:DiscMatterandGeo}

The general proofs in Sec. \ref{sec:GenProof} already show, that for a wide range of models that use geometrical clocks, polymerization and gauge fixing do not commute. However as shown in Sec. \ref{sec:gaugeconditions_Schwarzschild}, we can reverse engineer from a chosen lapse function (and shift vector) the corresponding gauge conditions. A difficulty lies in finding solutions of the corresponding PDE of the gauge fixing conditions which are similar in structure to (\ref{PDE2}). However, there still might exist inconsistencies when formulating the physical Hamiltonian. Going to models that use a mix of geometrical and matter clocks, we can already see in Sec. \ref{JorgeEd_model} that the situation simplifies. The geometrical part of the Hamiltonian
is not contributing to the stability equations of the gauge fixing condition. This also means that the solutions for lapse and shift vector of the stability equations will lose their connection dependency and thus any sensitivity to polymerization. This is the reason that matter clocks using unpolymerized matter, as considered in the previous subsection,  are able to bypass the polymerization.

Let us elaborate on this point by considering a reference model which incorporates spherically  symmetric Brown-Kucha\v{r} dust \cite{Brown:1994py} considered in the relational formalism in \cite{Giesel:2007wi} and its quantization within LQG in \cite{Giesel:2007wn}. After reducing the second class constraints, the system takes the following form \cite{Giesel:2007wi}
\begin{align}
\tensor*{C}{^{\text{dust}}} = \pm 4 \pi \pi_0 \sqrt{ \tensor{q}{^x^x} U_x U_x +1}, \quad \quad \tensor*{C}{^{\text{dust}}_x}= 4 \pi \qty(\pi_0 T_{,x} + \pi_x S_{,x}),
\end{align}
where the dust fields $\qty(T,S)$ have the canonical momenta $\qty(\tensor{\pi}{_0},\tensor{\pi}{_x})$, $U_x = - T_{,x} - \frac{\tensor{\pi}{_x}}{\tensor{\pi}{_0}} S_{,x}$ and the metric component is expressed by triads $\tensor{q}{^x^x}=\frac{E^a}{(E^b)^2}$.
Defining the matter clocks by the gauge fixing conditions $\tensor{G}{_0}= T - t$ and $\tensor{G}{_1}= S - x$, we can compute the solution of lapse function and shift vector to be
\begin{align}
N =\pm \frac{\sqrt{\tensor{q}{^x^x}\tensor*{\pi}{_x^2} + \tensor*{\pi}{_0^2}}}{\tensor{\pi}{_0}},  \quad \quad  N^x= - \tensor{q}{^x^x}\frac{\tensor{\pi}{_x}}{\tensor{\pi}{_0}}.
\end{align}
For such reference models (see Appendix. \ref{appendix} for further examples), the matter parts of the constraints only depend on the matter degrees of freedom as well as the  metric, thus the densitized triads. Since the gauge fixing conditions only depend on the matter degrees of freedom, no connection components occur in the stability equations and therefore in the lapse and the shift vector. This means that the stability equations are unchanged under polymerization and so the question whether polymerization and gauge fixing commute can be trivially answered for these models affirmatively.

We now consider the case where the matter sector is quantized with a loop quantization which yields a polymerization in the matter part at the effective level as well. It turns out that in this case gauge fixing and polymerization do no longer commute. To illustrate this let us discuss a polymerized version of the Brown-Kucha\v{r} dust model above (in the Appendix \ref{appendix} we have calculated this for the other reference models too). To incorporate matter polymerization, we define a general polymerization by replacing the canonical momentum of the matter fields by a general function
 \begin{equation}\label{polymerization_matter}
     \tensor{\pi}{_i} \mapsto h_i(\tensor{\pi}{_i};\lambda_i) \quad \quad \quad i = 0,x.
 \end{equation}
As mentioned earlier, it makes no difference for the stability equations in this model whether the geometric degrees of freedom are polymerized as well, so we can directly compute the stability of the gauge fixing conditions:
\begin{align}
\Dot{G_0} &\approx \qty(\pm N \frac{\tensor{h}{_0}}{\sqrt{\tensor{q}{^x^x}\tensor*{h}{_x^2}+ \tensor*{h}{_0^2}}})(\partial_{\pi_0} h_0)  -1 = 0 \quad \quad\Longrightarrow \quad N = \pm\frac{\sqrt{\tensor{q}{^x^x}\tensor*{h}{_x^2}+ \tensor*{h}{_0^2}}}{\tensor{h}{_0}(\partial_{\pi_0} h_0)},\\
\Dot{G_1} &\approx \qty(N^x \pm N \frac{\tensor{q}{^x^x}\tensor{h}{_x}}{\sqrt{\tensor{q}{^x^x}\tensor*{h}{_x^2} + \tensor*{h}{_0^2}}})(\partial_{\pi_x} h_x) = 0 \quad\,\,\,\,\,\Longrightarrow \quad N^x = -\frac{\tensor{q}{^x^x}\tensor{h}{_x}}{\tensor{h}{_0}(\partial_{\pi_0} h_0)}.
\end{align}
We can see, whenever the polymerization of the matter field $T$ is not the identity, we get a nontrivial contribution from the derivative of the corresponding polymerization function $h_0$ in the effective theory. This leads us to the conclusion that for models with matter clocks where the matter is polymerized, gauge fixing and polymerization do \textit{not} commute. This is not surprising, since in this case one is essentially treating the matter degrees of freedom in the same way as the geometrical ones under polymerization and so we are again in the scope of Lemma \ref{lemma}. One can see this in the second stability equation too. Note that here we can factor out the derivative term of $h_x$. This is possible since we do not have any temporal dependence of the gauge fixing condition. So in the partially gauge fixed system, where $G_0$ is fixed, gauge fixing and polymerization would again commute, a point we have discussed extensively in Sec. \ref{JorgeEd_model}. Furthermore above calculation is faithful example of Lemma \ref{corollary3} in action.

\section{Conclusions}
\label{sec:Conclusion}

In this manuscript, we have addressed the question of whether the procedure of gauge fixing and polymerization commutes for the effective dynamics of symmetry reduced models based on techniques of LQG. We have studied some representative situations in which the Gauss constraint is eliminated and  only the Hamiltonian constraint possibly along with the diffeomorphism constraint remains to be solved. To obtain dynamics in constrained systems one usually introduces gauge fixing conditions and their dynamical stability determines the corresponding Lagrange multipliers which are lapse and shift in the gravitational systems considered in this work. In practice for a given model one can also take the perspective of first choosing lapse and shift and then determining a set of gauge fixing conditions consistent with the dynamics. If we go beyond the classical theory and consider quantized models, then we can either implement the gauge fixing at the classical level or in the quantum theory, where the latter is treated at the effective level (polymerization) in this article. Therefore,  a pertinent question is whether polymerizing just the classical lapse and shift is sufficient to obtain solutions for the effective lapse and shift that are consistent with the stability of the effective gauge fixing conditions. In case  the preservation of the effective gauge fixing conditions with respect to the effective dynamics yields  solutions for lapse and shift which are simply the polymerized versions of the classical lapse and shift, then we say that gauge fixing and the polymerization as commutable with each other. If this is not the case, then polymerization and gauge fixing do not commute.

From our studies, we find that the commutativity of gauge fixing and polymerization depends, apart from the model under consideration, on the types of the gauge fixing conditions as well as their dependence on the temporal coordinate. For geometrical clocks, that we denote as the first type,  the gauge fixing condition involves only the geometrical degrees of freedom, whereas for matter clocks, denoted as the second type, they involve only matter degrees of freedom.  As  Lemma \ref{lemma} shows, when only the geometrical sector is polymerized and the gauge fixing conditions are of the first type with at least one of them depending on the temporal coordinate, then gauge fixing does not commute with polymerization. Lemma \ref{lemma} is then further illustrated by a simple but non-trivial example, i.e. the loop quantization of the interior of the Schwarzschild  black hole \cite{Corichi:2015xia}. In this example, due to the homogeneity of the Schwarzschild interior, the diffeomorphism constraint is fixed. As a result, the gauge fixing condition used to determine the lapse must depend on the temporal coordinate and thus Lemma \ref{lemma} applies to this setting. As a consequence, if we work with the same polymerization for the gauge fixing condition that is used for the Hamiltonian constraint, simply polymerizing the classical lapse in the effective theory leads to inconsistencies. Therefore, in practice, if we want to work with the lapse which is specified to yield well-known analytic solutions in the literature, we need to solve for the corresponding gauge fixing condition which obeys a PDE derived from consistency requirements. Our results reveal several properties of the gauge fixing conditions which hold  for both classical and effective dynamics. Firstly, as expected for a specific lapse, there can be multiple choices of the gauge fixing conditions and all of the applicable gauge fixing conditions are the solutions of the system. Secondly, if we choose a lapse for the effective dynamics that is the polymerization of the classical lapse, the corresponding consistent gauge fixing conditions for the effective dynamics are not the polymerization of their classical counterparts. Thirdly, although gauge fixing and polymerization do not commute in this case, the polymerized quantum system has the correct classical limit, that is to say, the effective Hamiltonian constraint, the effective lapse and the corresponding gauge fixing conditions can reduce to their classical counterparts when the polymerization parameters approach zero. It will be interesting to further investigate this issue for other loop  quantizations of the Schwarzschild interior, such as in \cite{Boehmer:2007ket,Olmedo:2017lvt,Ashtekar:2018cay, Bodendorfer:2019cyv} and understand the commutability of polymerization and gauge fixing conditions.

Relaxing and changing respectively some of the restrictions in the assumptions of Lemma \ref{lemma} can make gauge fixing commute with the polymerization. The first two corollaries of  Lemma \ref{lemma} highlight this.  In particular, when only the geometrical sector is polymerized, and if each of the gauge fixing conditions depend at most on one and in general different canononical pair and in addition none of the gauge fixing conditions depend on the temporal coordinate (Corollary \ref{lemma2}), or the time-dependent gauge fixing condition is of the second type (Corollary \ref{lemma3}), then gauge fixing turns out to commute with the polymerization. The gravitational collapse of an inhomogeneous dust cloud  which we have analyzed in Sec. \ref{JorgeEd_model} fits exactly into Corollary \ref{lemma3}. Since the time-dependent gauge fixing condition only depends on the dust field and thus is not affected by the polymerization of the geometric sector, the stability equations of the gauge fixing conditions for the effective dynamics are equivalent to those  obtained from the polymerization of the classical stability equations, leading to the solutions of the lapse and the shift which are the polymerization of their respective classical counterparts. It is important to note that with this type of the gauge fixing conditions, the choice of the lapse and shift for the effective dynamics should be exactly their polymerized classical counterparts otherwise inconsistencies in the dynamics would arise as is proved in Lemma \ref{lemma4}. We discuss some consequences of this in Sec. IIIB for the works in \cite{Kelly:2020lec, Kelly:2020uwj, Husain:2021ojz}. Our analysis shows pitfalls with some choices made in literature for the shift vector which are inconsistent with the stability of gauge fixing conditions.

Finally, we have also extended our analysis to the case in which the matter sector is polymerized and  matter clocks are chosen. In this particular case, which is an example of Lemma \ref{corollary3},  polymerization turns out to not commute with  gauge fixing for the same reason as in the case of choosing geometric clocks with a polymerized geometric sector addressed in Lemma \ref{lemma}. We choose the Brown-Kucha\v{r} dust model \cite{Brown:1994py,Giesel:2007wi} and the Gaussian dust model \cite{Kuchar:1990vy,Giesel:2012rb} as well as the four scalar fields model \cite{Giesel:2016gxq}  as examples in this category (in Appendix A). Due to the polymerization of the matter sector, the lapse and shift which are consistent with the effective dynamics include extra terms other than those obtained from the polymerization of the momenta of the matter fields. In general our analysis shows that whenever the gauge fixing condition that involves a temporal dependence involves variables that belong to a set of canonical variables where at least one of the elementary variables is polymerized we are in the situation that gauge fixing and polymerization will not commute. As a result, in order to deal with a model where gauge fixing and polymerization commute, one could for instance choose matter clocks for the Hamiltonian constraint and only polymerize the geometric degrees of freedom or choose geometric clocks for the unpolymerized Hamiltonian constraint and apply a polymerization to the matter sector only. From the LQG perspective the latter will be a rather unnatural choice.

In summary, our examples illustrate that when polymerizing a classical system with gauge degrees of freedom,  gauge fixing and polymerization may not commute with one another in general. However, this causes no issue a priori because independent from the classical theory the requirement of the stability of the gauge fixing conditions yields a clear procedure how the effective lapse and shift that are consistent with the effective dynamics can be determined for a given choice of gauge fixing conditions. This allows to check for each given model in practice whether gauge fixing and polymerization commute. If one wants to work with a specific choice of a polymerized lapse and/or shift then one can look for possible gauge fixing conditions that yield this choice and that are consistent with the effective dynamics. As shown in our work depending on the model this might require to use different polymerizations for constraints and gauge-fixing conditions and/or work with polymerized gauge fixing conditions that do not correspond to a polymerization of the classical gauge-fixing condition. These are undesirable features which should be avoided for a well motivated and consistent polymerized model.

As a further remark, compared with other requirements on the effective dynamics,  such as  a specific choice for lapse and/or shift motivated from the effective constraint algebra, we believe the first priority should be placed on the consistency in the dynamics generated by the effective Hamiltonian, that is, the gauge fixing conditions must be preserved by the effective dynamics and one has to carefully check whether choices one motivates from the constraint algebra do not violate this consistency requirement. Under this necessary condition, the commutativity/non-commutativity of the gauge fixing and the polymerization depends on the types of gauge fixing conditions and also on which sector of the system is polymerized.

As a final remark, next to performing the gauge fixing in the effective theory, that corresponds to applying a gauge fixing in the quantum theory, one can also apply the relational formalism and the construction of Dirac observables in the quantum theory. Although a detailed analysis on this topic goes beyond the scope of this paper our results give some first insight on the relation between classical and quantum clocks for those classes of gauge fixing conditions where we can relate the gauge-fixed theory to the reduced phase space of elementary Dirac observables and their dynamics. As our results show, if one quantizes the gauge fixing  condition with the same polymerization as used for the constraints, that is what one would usually do, then only in very special cases will a dynamically consistent lapse and shift be given by the polymerization of their classical counterparts. In the generic case this means that choosing a classical or quantum clock has an effect on how the quantization of lapse and shift needs to be performed.
Moreover, our analysis also shows that whether such a commutativity exists strongly depends on how the variables and their conjugates involved in the temporal gauge fixing condition are quantized, which at an effective level means whether the variables and/or their conjugate variables are polymerized. Among other criteria such as that the algebra of Dirac observables has a simple structure and the physical Hamiltonian can be implemented as a well defined operator on the physical Hilbert space, the requirement of commutativity of quantization and gauge fixing in the sense discussed in this work can allow to further discriminate possible different choices of clocks in the relational formalism.

\section*{Acknowledgments}

This work is supported by the DFG-NSF grants PHY-1912274 and 425333893.

\appendix
\section{The matter clocks  with polymerized matter sector in the Gaussian dust and the four scalar reference field models}
\renewcommand{\theequation}{A.\arabic{equation}} \setcounter{equation}{0}
\label{appendix}
In this appendix, we extend the analysis of reference matter models. We will start with the Gaussian dust model discussed in \cite{Giesel:2012rb, Kuchar:1990vy}. As before we do not specify to any symmetry reduced model yet but keep our analysis as general as possible and formulate the effective model by polymerizing a suitable subset of the canonical variables. The Gaussian dust gives rise to the following additional terms to the geometrical parts of the constraints \cite{Giesel:2012rb}
\begin{align}
    \tensor{C}{^{\text{dust}}} &= \pm\qty(\tensor{p}{_0}\sqrt{1+\tensor{q}{^a^b}\tensor{T}{_{,a}}\tensor{T}{_{,b}} } + \frac{\tensor{q}{^a^b}\tensor{T}{_{,a}} \tensor{\pi}{_j} \tensor*{S}{^j_{,b}} }{\sqrt{1+\tensor{q}{^a^b}\tensor{T}{_{,a}}\tensor{T}{_{,b}} }}),\\
    \tensor*{C}{^{\text{dust}}_x} &= \tensor{\pi}{_0}\tensor{T}{_{,a}} + \tensor{\pi}{_j}\tensor*{S}{^j_{,a}},
\end{align}
where the four dust fields $\qty(T,\tensor*{S}{^j})$ have the canonical momenta $\qty(\tensor{\pi}{_0},\tensor{\pi}{_j})$. To construct the reference model we will use these fields as clocks, i.e. $\tensor{G}{_0}= T - t$ and $\tensor{G}{_j}= \tensor*{S}{^j} -\tensor*{x}{^j}$. We will directly compute the stability of these gauge fixing conditions in the effective theory using the polymerization of the matter degrees of freedom like in equation (\ref{polymerization_matter}). Note that we can go back to the classical theory by setting $h_i(\tensor{\pi}{_i};\lambda_i)\rightarrow \tensor{\pi}{_i}$. Straightforward calculations show that
\begin{align}
    \Dot{G_0} &=\pm N(\partial_{\pi_0} h_0)\sqrt{1+\tensor{q}{^a^b}\tensor{T}{_{,a}}\tensor{T}{_{,b}} }  -1 \approx \pm N(\partial_{\pi_0} h_0)  -1 = 0 \quad \quad\Longrightarrow \quad N = \pm\frac{1}{\partial_{\pi_0} h_0},\\
    \Dot{G_j} &= N^a\tensor*{S}{^j_{,a}} \pm N \frac{\tensor{q}{^a^b}\tensor{T}{_{,a}}  \tensor*{S}{^j_{,b}} }{\sqrt{1+\tensor{q}{^a^b}\tensor{T}{_{,a}}\tensor{T}{_{,b}} }} \approx N^a\tensor*{S}{^j_{,a}} = 0 \quad\,\,\,\,\,\Longrightarrow \quad N^a = 0.
\end{align}
Similar to the  Brown-Kucha\v{r} dust model \cite{Brown:1994py,Giesel:2007wi,Giesel:2007wn} discussed in Sec. \ref{sec:DiscMatterandGeo}, we can see that once again polymerization and gauge fixing do not commute. To be precise, in the classical theory the lapse equals unity. Switching to the effective theory this is not the case anymore, since lapse is now the inverse derivative of the polymerization of the temporal reference field. These two solutions only coincide for constant polymerizations, which are trivial and thus will be excluded. In both cases the shift vector is vanishing.

Another reference model in the literature is the four Klein-Gordon scalar field model of \cite{Giesel:2016gxq}. We will again treat this model in the symmetry unrestricted sector. Although this is not a dust model, we will adapt the notation of the cited paper to our notation used for matter models. The matter part constraints take the form \cite{Giesel:2016gxq}
\begin{align}
    \tensor{C}{^{\text{dust}}} &= \frac{\tensor*{\pi}{_0^2}}{2 \sqrt{q}} + \frac{1}{2} \sqrt{q}\,\tensor{q}{^a^b}\tensor{T}{_{,a}}\tensor{T}{_{,b}} - \sum_j \tensor{\pi}{_j} \sqrt{\tensor{q}{^a^b} \tensor*{S}{^j_{,a}}\tensor*{S}{^j_{,b}} },\\
    \tensor*{C}{^{\text{dust}}_x} &= \tensor{\pi}{_0}\tensor{T}{_{,a}} + \tensor{\pi}{_j}\tensor*{S}{^j_{,a}},
\end{align}
where we used the abbreviation $\sqrt{q}\coloneqq\sqrt{\abs{\det(\tensor{q}{^a^b})}}$. Using the same gauge fixing conditions as before we compute
\begin{align}
    \Dot{G_0} &= N(\partial_{\pi_0} h_0)\frac{\tensor*{h}{_0}}{\sqrt{q}}  -1 = 0 \quad \quad\Longrightarrow \quad N = \frac{\sqrt{q}}{\tensor*{h}{_0}\qty(\partial_{\pi_0} h_0)},\\
    \Dot{G_j} &= \qty(\partial_{\pi_j} h_j)\qty(N \sqrt{\tensor{q}{^a^b} \tensor*{S}{^j_{,a}}\tensor*{S}{^j_{,b}} } + N^a\tensor*{S}{^j_{,a}}) = 0 \quad\,\,\,\,\,\Longrightarrow \quad N^a = N \sqrt{\tensor{q}{^a^a}}.
\end{align}
Similar to the reference model analyzed above, a partial derivative of the polymerization is appearing. This breaks the commutativity of gauge fixing and polymerization.

\end{document}